\pdfoutput=1
\documentclass[structabstract, longauth]{aa}  
\usepackage{graphicx}
\usepackage{subfigure}
\usepackage{txfonts}
\usepackage{natbib}
\newcommand{\msun}{\mathrm{M}_\odot}
\bibpunct{(}{)}{;}{a}{}{,}
\titlerunning{A new galaxy classification based on the global optical properties of zCOSMOS 10k sample.}
\authorrunning{G. Coppa et al.}
\begin{document}
   \title{The bimodality of the 10k zCOSMOS--bright galaxies up to $z\sim 1$:
a new statistical and portable classification based on the global optical galaxy properties}

   \subtitle{}

   \author{G. Coppa\inst{1,2}
          \and
          M. Mignoli\inst{1}
          \and
          G. Zamorani\inst{1}
          \and
          S. Bardelli\inst{1}
          \and
          M. Bolzonella\inst{1}
          \and
          L. Pozzetti\inst{1}
          \and
          D. Vergani\inst{1,3}
          \and
          E. Zucca\inst{1}
          \and
          A. Cimatti\inst{2}
          \and
          S.~J. Lilly\inst{4}
          \and
          C.~M. Carollo\inst{4}
          \and
          T. Contini\inst{5}
          \and
          O. Le~F\`evre\inst{6}
          \and
          A. Renzini\inst{7}
          \and
	  M. Scodeggio\inst{3}
          \and
          V. Mainieri\inst{8}
          \and
          A. Bongiorno\inst{9}
          \and
          K. Caputi\inst{4}
          \and
          O. Cucciati\inst{6,10}
          \and
          S. de~la~Torre\inst{6,10,3}
          \and
          L. de~Ravel\inst{6}
          \and
          P. Franzetti\inst{3}
          \and
          B. Garilli\inst{3}
          \and
          P. Memeo\inst{3}
          \and
          A. Iovino\inst{10}
          \and
          P. Kampczyk\inst{4}
          \and
          J.-P. Kneib\inst{6}
          \and
          C. Knobel\inst{4}
          \and
	  A. M. Koekemoer\inst{17}
	  \and
          K. Kova\v c\inst{4}
          \and
          F. Lamareille\inst{2,5}
          \and
          J.-F. le~Borgne\inst{5}
          \and
          V. le~Brun\inst{6}
          \and
          C. Maier\inst{4}
          \and
          R. Pell\`o\inst{5}
          \and
          Y. Peng\inst{4}
          \and
          E. Perez-Montero\inst{5}
          \and
          E. Ricciardelli\inst{7}
          \and
          C. Scarlata\inst{4}
          \and
          J.~D. Silverman\inst{4,8}
          \and
          M. Tanaka\inst{8}
          \and
          L. Tasca\inst{6,3}
          \and
          L. Tresse\inst{6}
          \and
          U. Abbas\inst{6}
          \and
          D. Bottini\inst{3}
          \and
          P. Capak\inst{11,12}
          \and
          A. Cappi\inst{2}
          \and
          P. Cassata\inst{6}
          \and
          M. Fumana\inst{3}
          \and
          L. Guzzo\inst{10}
          \and
          A. Leauthaud\inst{6}
          \and
          D. Maccagni\inst{3}
          \and
          C. Marinoni\inst{13}
          \and
          B. Meneux\inst{9,14}
          \and
          P. Oesch\inst{4}
          \and
          C. Porciani\inst{4}
          \and
          R. Scaramella\inst{16}
          \and
          N. Scoville\inst{11}
          }

   \institute{Dipartimento di Astronomia, Universit\`a di Bologna, Via Ranzani 1, I-40138, Bologna\\
              \email{graziano.coppa@studio.unibo.it}
         			\and
              INAF -- Osservatorio Astronomico di Bologna, Bologna, Italy
	     	      \and
	     	 		  INAF -- Istituto di Astrofisica Spaziale e Fisica Cosmica, Milano, Italy
 				      \and
 				      Institute of Astronomy, ETH Z\"urich, Z\"urich, Switzerland
 				      \and
 				      Laboratoire d'Astrophysique de Toulouse-Tarbes, Universit\'e de Toulouse, CNRS Toulouse, France
 				      \and
 				      Laboratoire d'Astrophysique de Marseille, Marseille, France
 				      \and
 				      Dipartimento di Astronomia, Universit\`a di Padova, Padova, Italy
 				      \and
 				      European Southern Observatory, Garching, Germany
 				      \and
 				      Max Planck Institut f\"ur Extraterrestrische Physik, Garching, Germany
 				      \and
 				      INAF -- Osservatorio Astronomico di Brera, Milano, Italy
 				      \and
 				      California Institute of Technology, Pasadena CA, USA
 				      \and
 				      Spitzer Science Center, Pasadena CA, USA 			
 				      \and
 				      Centre de Physique Theorique, Marseille, France	 
 				      \and
 				      Universitats-Sternwarte, M\"unchen  
 				      \and
 				      INAF -- Osservatorio Astronomico di Arcetri, Firenze, Italy
 				      \and
 				      INAF -- Osservatorio Astronomico di Roma, Monte Porzio Catone, Italy
			              \and
				      Space Telescope Science Institute, Baltimore, USA}

   \date{Preprint online version: September 3rd, 2010}

 
  \abstract
   {}
   {Our goal is to develop a new and reliable statistical method to classify galaxies from large surveys. We probe the reliability of the method by comparing it with a three-dimensional classification cube (Mignoli et al.~2009), using the same set of spectral, photometric and morphological parameters.}
   {We applied two different methods of classification to a sample of galaxies extracted from the zCOSMOS redshift survey, in the redshift range $0.5 \la z \la 1.3$. The first method is the combination of three independent classification schemes -- a spectroscopic one based on the strength of the continuum break at 4000 \AA\ and the rest-frame equivalent width of [\ion{O}{ii}] emission line, a photometric one based on observed $B-\mathrm{z}$ colours, a morphological one adapted from Scarlata et al.~(2009) --, while the second method exploits an entirely new approach based on statistical analyses like Principal Component Analysis (PCA) and Unsupervised Fuzzy Partition (UFP) clustering method. The PCA+UFP method has been applied also to a lower redshift sample ($z \la 0.5$), exploiting the same set of data but the spectral ones, replaced by the equivalent width of H$\alpha$.}
   {The comparison between the two methods shows fairly good agreement on the definition on the two main clusters, the \emph{early-type} and the \emph{late-type} galaxies ones. Our PCA-UFP method of classification is robust, flexible and capable of identifying the two main populations of galaxies as well as the intermediate population. The intermediate galaxy population shows many of the properties of the ``green valley'' galaxies, and constitutes a more coherent and homogeneous population. The fairly large redshift range of the studied sample allows us to behold the \emph{downsizing} effect: galaxies with masses of the order of $3\cdot 10^{10}\, \msun$ mainly are found in transition from the late type to the early type group at $z>0.5$, while galaxies with lower masses -- of the order of $10^{10}\, \msun$ -- are in transition at later epochs; galaxies with $M <10^{10}\, \msun$ did not begin their transition yet, while galaxies with very large masses ($M > 5\cdot 10^{10}\, \msun$) mostly completed their transition before $z\sim 1$.} 
   {}


   \maketitle
%

\section{Introduction}

It is well known that galaxies show a large assortment of observational and intrinsic features. In the local and near universe, \citep[up to $z \sim 1$,][]{Bell} many of these properties, such as optical colours \citep{Strateva, Ball}, morphological parameters \citep{Driver}, and spectral indices \citep{Kauffmann03, Balogh}, are known to come in a bimodal fashion. The origin of these bimodalities is not clear yet, in terms of galaxy evolution \citep{Blanton}. The existence of two different groups has been explained in the past as a matter of different initial conditions (galaxies having different mechanisms of formation), whether it would be a dissipationless collapse, leading to the formation of an elliptical galaxy and the dispersion of its gas content, or a dissipative one, giving as a result a spiral galaxy which retained its gas and could subsequently maintain its star formation \citep{Ellis}. The most accepted current cosmological models, however, predict that the formation of galaxies is mostly hierarchical, massive ellipticals being the result of a series of major mergers between smaller spiral galaxies \citep[][for a review]{Cole, Baugh, Schweizer}. For these reasons the widely accepted scenario to explain the bimodal segregation of the galaxy properties is an evolutive one: galaxies in different phases of their evolution show different colours, different star formation rates, different morphologies. How these different parameters are connected is still a matter of debate \citep{Conselice06}; it appears clear, however, that a better knowledge of these connections would help develop a deeper understanding of the physical processes behind galaxy evolution.

The purpose of this work is to develop a robust and powerful method to classify galaxies from large surveys, in order to establish and confirm the connections between the principal observational parameters of the galaxies (spectral features, colours, morphological indices), and help unveil the evolutions of these parameters from $z \sim 1$ to the local Universe. This paper makes use of zCOSMOS and COSMOS surveys data, and capitalizes their large capabilities in terms of data reliability and vastness.

The paper is organized as follows: in \S\ref{sec:zcosmos} we will briefly describe the zCOSMOS survey and the sub-samples of the data used in this paper; in \S\ref{sec:cube} we will present the extension to the 10k zCOSMOS-bright sample of the classification cube method presented by \citet[hereafter M09]{Mignoli09} as applied to a smaller sample; in \S\ref{sec:PCAclust} we will present a new method of classification, based on statistical tools like Principal Component Analysis and Cluster Analysis; in \S\ref{sec:results} we will discuss and comment results of the two combined methods, and present a quick review of some interesting sub-populations; in \S\ref{sec:conclusioni} we will present final remarks and the general picture emerging from this work. 

Throughout this paper, unless otherwise stated, we assume a concordance cosmology with $\Omega_{\mathrm{M}} = 0.25$, $\Omega_{\Lambda} = 0.75$ and $H_{0} = 70$ km s$^{-1}$ Mpc$^{-1}$; magnitudes are expressed in the AB system.


\section{Description of zCOSMOS}\label{sec:zcosmos}

zCOSMOS \citep{Lilly07, Lilly09} is a large redshift survey which has been carried out using VIMOS spectrograph \citep{LeFevre} installed at the 8 m UT3 ``Melipal'' of the European Southern Observatory's Very Large Telescope at Cerro Paranal. The main goal of the survey is to trace the large scale structure of the universe up to $z \sim 3$ and to characterize galaxy groups and clusters. 

In order to exploit more efficiently the resources of the VIMOS spectrograph, the zCOSMOS survey has been split in two distinct parts:
\begin{itemize}
 \item zCOSMOS-bright, a magnitude-limited ($I_{\mathrm{AB}} < 22.5$) survey that, once completed, will consist of $\sim 20\,000$ galaxies in a redshift range of $0.1 < z < 1.2$. This part of the survey is being undertaken on the 1.7 deg$^2$ COSMOS field fully covered by the ACS camera of the Hubble Space Telescope \citep{Koekemoer07};
 \item zCOSMOS-deep, a survey whose $\sim 10\,000$ galaxies are selected through various colour criteria, with a redshift range of $1.4 < z < 3.0$, in the central 1 deg$^2$ of the COSMOS field.
\end{itemize}

The specifications of the bright part of the survey include a very high success rate in redshift determination ($\sim 90\%$), a uniform sampling rate across the whole field, and fairly good velocity accuracy ($\sim 100 \mbox{ km s}^{-1}$) which allow to define the dynamical environment of the galaxies.



The data release this paper is based upon, called \emph{10k sample}, is made up of 10\,642 galaxies from the zCOSMOS-bright part of the survey, regardless of the spectral quality. Our first work sample is composed by 4\,874 galaxies between $0.48 < z < 1.28$: this will be referred to as \emph{high redshift whole sample}. This choice is due to the fact that, given the spectral range of the observations (5550-9650 \AA), the spectral features around rest-frame 4000 \AA\ that we use in this work (the continuum break at $\sim 4000$ \AA\ -- from now on $D4000$ -- and the [\ion{O}{ii}] emission line) can be detected only in that redshift range. The \emph{high redshift high quality sample}, instead, is composed by all the galaxies with spectroscopic flag 4, 3 and 2.5, i.e. galaxies with secure redshifts, or likely redshifts confirmed by the photometric one \citep[for a more detailed review of spectral confidence flags, see][]{Lilly09}. Galaxies with spectroscopic flag=1 are excluded because of their poorly-defined spectral features, while flag=9 are excluded because of the absence of other spectral features beside a single strong emission line; this high quality subset is composed by 3\,720 objects (76\% of the whole sample). The subsequent extension of the work to lower redshifts, achieved by substituting $D4000$ and $EW_0[\ion{O}{ii}]$ with the rest-frame equivalent width of H$\alpha$ ($EW_0(\mathrm{H}\alpha)$), builds up a different dataset composed by 3\,402 galaxies (\emph{low redshift whole sample}); the corresponding \emph{low redshift high quality sample} is made up by 3\,005 galaxies (88\% of the whole sample). It has to be noted that, throughout the analysis, the informations on the errors associated with the parameters were not included, since many parameters (like the morphological ones) were not given an error. Furthermore, spectroscopical stars and broad-line active galactic nuclei have been excluded from both samples.

\section{The classification cube method}\label{sec:cube}

We extended the classification method developed by M09, applied to the first release of the zCOSMOS-bright catalogue (the so-called \emph{1k sample}, composed by $\sim 1\,000$ galaxies) to the larger dataset provided by the 10k sample. This classification is based on three independent datasets (spectroscopic, photometric, morphological) which exploit the bimodality shown by galaxies in many features.

\subsection{Spectral classification}\label{subsec:spectral}

Spectral measurements of the 10k sample were carried out by the automatic computer code \verb|PlateFit| \citep{Lamareille}. The program analyses the galaxy spectra and performs measurements of equivalent width and flux for the most important spectral features.

We classified galaxies in the sample using the diagram $D4000$ vs.~rest-frame equivalent width of [\ion{O}{ii}] (from now on $EW_0[\ion{O}{ii}]$) developed by \citet{Cimatti} and extensively used in many works, e.g. \citet{Kauffmann04, Mignoli05, Franzetti}. $D4000$ is a tracer of cumulative star formation: galaxies with stronger 4000~\AA\ breaks had a longer history of forming stars \citep{Bruzual83, Marcillac}; on the other hand, the presence of [\ion{O}{ii}] in emission is an effective signature of ongoing star formation \citep{Kewley, Kennicutt98}. Upper limits to the observed equivalent widths of [\ion{O}{ii}] emission lines have been computed using the empirical relation proposed by \citet{Mignoli05}, and compared to the values of the upper limits produced by \verb|PlateFit|. The empirical envelope relation, which replaces \verb|PlateFit| upper limits when those are lower, is:

\begin{equation}\label{eq:envelope}
EW_{\mathrm{lim}} = \frac{SL \cdot \Delta}{S/N_{\mathrm{cont}}}
\end{equation}

where $SL=3$ is the significance level of each line
, $\Delta$ is the spectrum resolution (in \AA) and $S/N_{\mathrm{cont}}$ is the signal-to-noise ratio of the spectrum calculated in the proximity of the line.

\begin{figure}[t]
\begin{center}
\includegraphics[width=.45\textwidth]{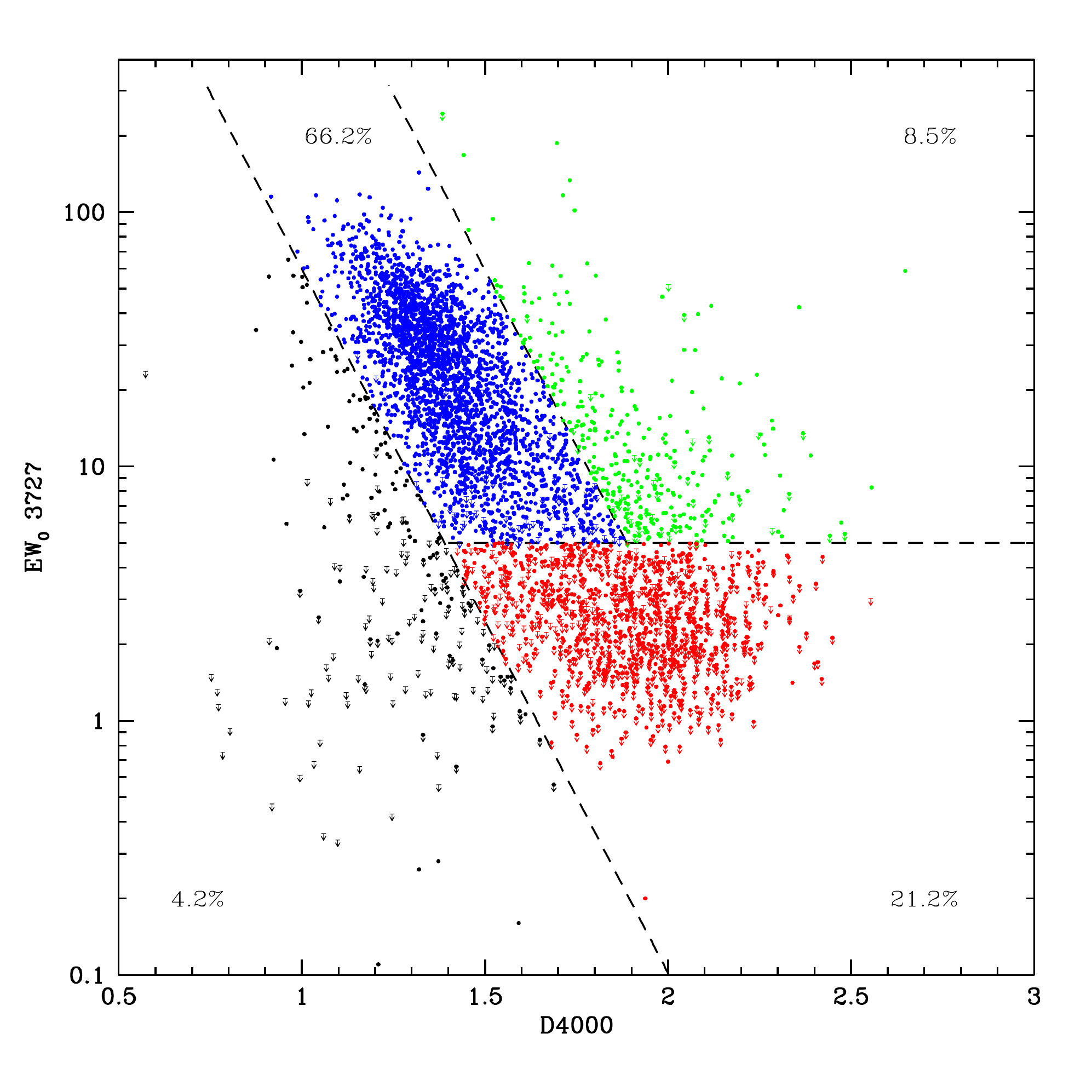}
\caption{Spectral classification diagram for the 10k high quality zCOSMOS sample. In red are \emph{passive galaxies}, in blue \emph{star forming galaxies} and in green \emph{red emitters}. Small arrows mark objects for which we have only upper limits in $EW_0[\ion{O}{II}]$. Numbers represent the fraction of objects belonging to each class.}\label{fig:specclass}
\end{center}
\end{figure}

In Fig.~\ref{fig:specclass} the $D4000$-$EW_0[\ion{O}{ii}]$ plane is shown. The horizontal dashed line represents the cut at 5~\AA\ in $EW_0[\ion{O}{ii}]$ used to separate strong and weak line emitters, adopted by M09. We used an iterative $\sigma$-clipping least squares algorithm to constrain the regions of highest density obtaining the following boundaries:\\
\begin{equation}
 1.64 \leq D4000 + 0.36 \log(EW_0[\ion{O}{ii}]) \leq 2.14
\end{equation}

This is somewhat narrower with respect to Eq.~(2) in M09, especially toward the left side of the diagram -- low $D4000$ values -- due to a lower $\sigma$ rejection in the algorithm.

We defined star-forming galaxies the 66\% of the spectroscopic high quality galaxies with low values of $D4000$ and high values of $EW_0[\ion{O}{ii}]$, and quiescent galaxies (21\%) those with low values of $EW_0[\ion{O}{ii}]$ and high values of $D4000$. Galaxies populating the upper-right part of the diagram, which are the 8.5\% of the total, are defined as the population of intermediate galaxies with a quiescent-like continuum but with strong emission lines, and are mainly associated with AGNs.

The left part of the diagram is mainly populated by low quality spectra objects; high quality objects in this region (which are 4\% of the total high quality sample) reside mostly near the boundary.

Considering the high quality sample only, nearly 88\% of the galaxies are classified in one of the two main classes. Relaxing the constraints on the requested confidence on the spectral features, the fraction of galaxies in each area of the $D4000$-$EW_0[\ion{O}{ii}]$ plane is mostly unchanged.


\subsection{Photometric classification}\label{subsec:photo}

\begin{figure}[t]
 \begin{center}
 \includegraphics[width=.45\textwidth]{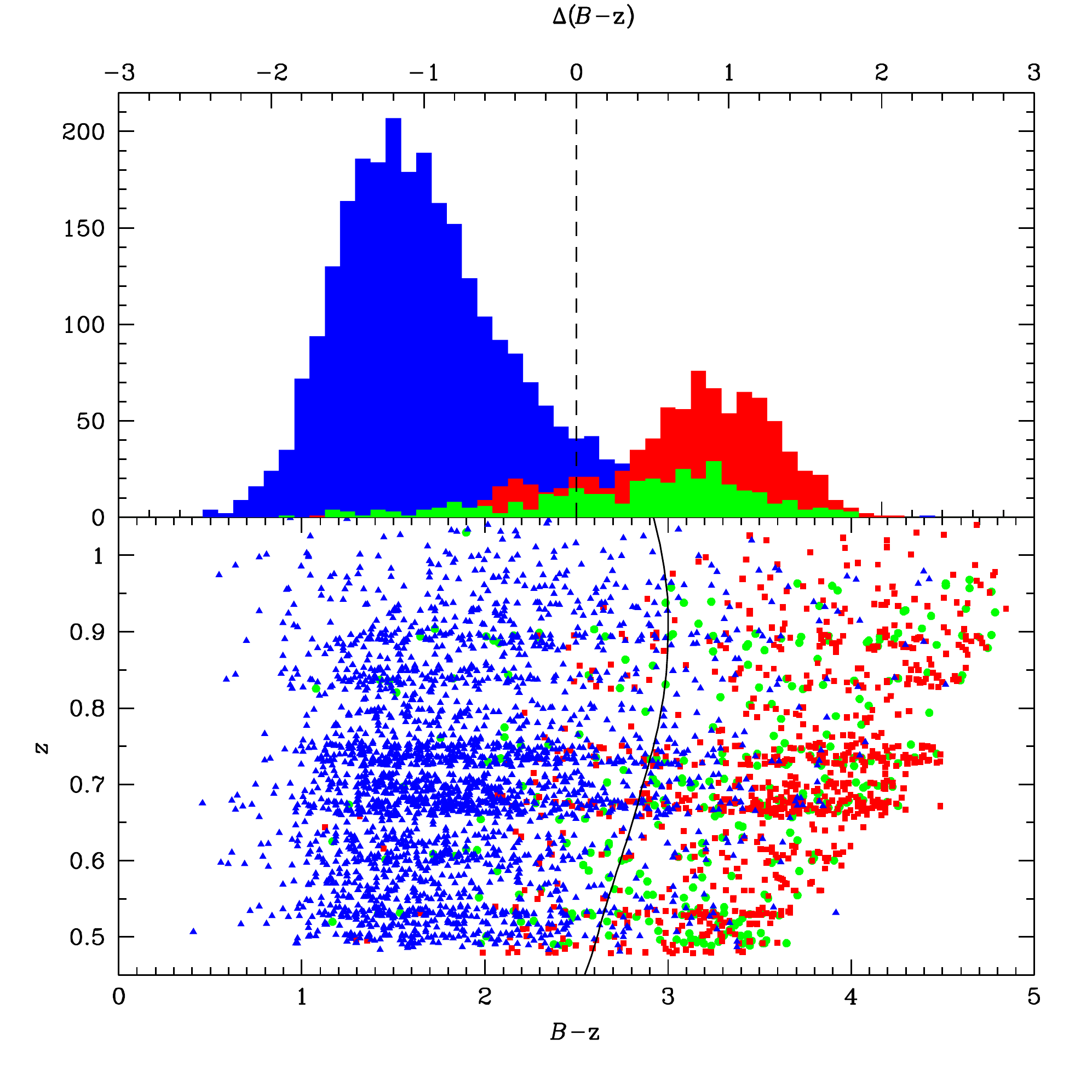}
\caption{Photometric classification of the 10k zCOSMOS-bright high quality sample. In the lower panel colour $B-\mathrm{z}$ versus redshift $z$ is shown: blue triangles are star-forming, red squares are quiescent, green dots are red emitting galaxies. Solid line represents the evolutionary $B-\mathrm{z}$ track of a template Sab galaxy from \citet{Coleman} \citep{Sawicki}. In the upper panel the distributions of $\Delta (B-\mathrm{z})$, as defined in Eq.~(\ref{eq:delta}), for star-forming galaxies (blue histogram), quiescent galaxies (red histogram) and red emitting galaxies (green histogram) are plotted. The dashed line represents $\Delta (B-\mathrm{z})$ of the Sab galaxy evolutionary track used as separator.}\label{fig:photo}
\end{center}
\end{figure}

We introduce another classification based on the photometric properties of the galaxies. In the lower panel of Fig.~\ref{fig:photo} the colour $B-\mathrm{z}$ of the galaxies \citep{Capak} is shown as a function of their redshift. We used $B-\mathrm{z}$ colour because of its effectiveness in separating the two galaxy classes in the redshift range explored by the zCOSMOS bright sample (M09). Spectroscopic star-forming galaxies (blue triangles) have lower $B-\mathrm{z}$ and thus are bluer than both quiescent and intermediate galaxies (respectively red squares and magenta dots). As a way of discriminating the two populations, we used the colour track of a Sab galaxy template, from the set provided by \citet*{Coleman} (see discussion in M09).


Galaxies classified as intermediate on the basis of their spectral properties are distributed in the same region as the quiescent ones; this can be seen in the upper panel, where is plotted the distribution of the distances between measured colours and the colour of the template at the redshift of the galaxy:
\begin{equation}\label{eq:delta}
 \Delta (B-\mathrm{z}) = (B-\mathrm{z})_{\mathrm{obs}} - (B-\mathrm{z})_{\mathrm{templ}}
\end{equation}

We use the quantity $\Delta (B-\mathrm{z})$ to segregate photometrically the galaxies: if $\Delta (B-\mathrm{z}) > 0$ galaxies are considered ``red'', while when $\Delta (B-\mathrm{z}) < 0$ galaxies are put in the ``blue'' class. Since, as we said, intermediate galaxies seem to share colours with the quiescent galaxies, we decided to merge these spectroscopic classes into one general ``quiescent'' category.

\begin{table}[tb]
\begin{center}
  \begin{tabular}{cccc}
 \hline
  $B-\mathrm{z}$ & Quiescent & Star-forming & Total \\
 \hline
  Red & 983 (1167) & 227 (318) & 1210 (1485)\\
  Blue & 208 (320) & 2431 (3081) & 2639 (3401)\\
  Total & 1191 (1487) & 2658 (3399) & 3849 (4886)\\
\hline
\end{tabular}
\caption{Summary of the number of high spectral quality galaxies in spectroscopic and photometric classifications. Between parentheses are figures from the whole sample.}\label{tab:specphoto}
\end{center}
\end{table} 

In Table \ref{tab:specphoto} the 2x2 contingency table for spectral and photometric classifications is shown: almost 90\% of the high quality sample shows a full agreement between the spectral and photometric classifications (87\% for the whole sample). The Cohen's kappa coefficient for inter-rater agreement is 0.74, confirming that the classifications are statistically consistent.

\subsection{Morphological classification}

Morphology data are provided by \citet{Scarlata}, who built their \emph{Zurich Estimator of Structural Types} (ZEST) performing a Principal Component Analysis (PCA) on 5 parameters derived directly from HST/ACS images of the COSMOS survey \citep{Koekemoer07}.


The ZEST classification scheme adopts a main morphological index, which is 1 (for elliptical galaxies), 2 (for spirals) or 3 (for irregulars), plus an integrative bulgeness parameter (only for galaxies with main index of 2), calculated from galaxy S\'ersic indexes. In this way spiral galaxies are further divided into four subclasses: 2.0, 2.1, 2.2, 2.3 going from bulge dominated spirals to disk dominated, largely following Hubble classification of spiral galaxies from S0 through Sc types.




We assigned ZEST type 2.2, 2.3 and 3 galaxies to a common morphological category, the \emph{disk-dominated and irregular galaxies}, and
ZEST types 1 and 2.0 to 
another common category, the 
\emph{ellipsoidal galaxies}. ZEST types 2.1 (spiral galaxies with an intermediate bulge-to-disk ratio) are 
furtherly divided according to their colour properties:
\begin{table}[b]
\begin{center}
  \begin{tabular}{cccc}
 \hline
  ZEST $\backslash$ spectral & Quiescent & Star-forming & Total \\
 \hline
  ellipsoidal 	& 607 (717) & 236 (292)  & 843 (1009) \\
  2.1 		& 350 (410) & 436 (528) & 786 (938)\\
  disk-dominated&   141 (232)  &  1860 (2391) & 2001 (2623)  \\ 
\hline
  Total  	 & 1098 (1359) & 2532 (3211)  &  3630 (4570)\\
\hline
\end{tabular}
\caption{Summary of the number of high spectral quality galaxies in spectroscopic and photometric classifications. Between parentheses are figures from the whole sample.}\label{tab:specmorpho}
\end{center}
\end{table} 
indeed, most (83\%, $360/436$) spectroscopic star-forming galaxies of ZEST type 2.1 have a negative $\Delta (B-\mathrm{z})$, and are therefore classified as ``blue'', while a similar percentage (82\%, $287/350$) of spectroscopic quiescent galaxies have $\Delta (B-\mathrm{z}) > 0$ and are classified as ``red''. Therefore, we included the ``red'' population of the ZEST 2.1 type in the morphologically ellipsoidal class and the ``blue'' population of them in the disk-dominated class (see discussion in M09).

In Table~\ref{tab:specmorphodef}, we present the numerical results of our morphological classification. The Cohen's kappa coefficient is $\approx 0.67$ for the high quality sample, proving the goodness of our classifications.

\begin{table}[b]
\begin{center}
 \begin{tabular}{cccc}
 \hline
  morph $ + B-\mathrm{z}$ & Quiescent & Star-forming & Total \\
 \hline
  Spheroidal 	& 894 (1049) & 312 (394) & 1206 (1443) \\
  Disk/Irregular& 204 (310) & 2220 (2817)  & 2424  (3127) \\
 \hline
  Total 	& 1098 (1359) & 2532 (3211) & 3630  (4570)\\
\hline
\end{tabular}
\caption{Spectral--morphological contingency table. Figures are for the high quality sample; between parentheses are figures for the full sample.}\label{tab:specmorphodef}
\end{center}
\end{table} 

\subsection{The cube}

To better analyse the correlations and similarities of our galaxies, we merged the three classifications (spectroscopic, photometric and morphological) into a three-axial framework, a \emph{classification cube}. To simplify the classification we assigned to each galaxy a 3-digit numerical flag which encompasses information from the three categories:

\begin{itemize}
 \item The first digit represents the spectral classification. Flags 1 and 2 classify a galaxy as a ``quiescent'' and ``star-forming'' type, respectively.
 \item the second digit stands for the colour classification. Flag 1 and 2 classify a galaxy as a ``red'' and ``blue'' type, respectively.
 \item the third digit is the morphological flag. Flags 1 and 2 classify a galaxy as a ``spheroidal'' and ``disk/irregular'' type, respectively.
\end{itemize}

So, for instance, a ``212'' classificator denotes a star-forming, disk-dominated galaxy with $\Delta (B-\mathrm{z}) > 0 $, therefore red.

Table~\ref{tab:cube} shows the summary of the 3D classification cube. Removing from the \emph{high redshift whole sample} objects for which the full set of data was not available, the full sample of the cube retains 4\,600 sources, while the high quality sub-sample is made up of 80\% of them (3\,630). Figures change very little between the two samples: almost 60\% of the sources show a fully concordant ``222'' classification (star-forming spectra, blue colours, disk-dominated morphologies) and more than 20\% of the sample is composed by ``111'' galaxies (quiescent spectra, red colours, spheroidal morphologies). On the whole, 83\% of the galaxies show a fully concordant cube classification, very similar to the 85\% of concordance shown by the smaller zCOSMOS-bright 1k sample (see M09).

\begin{table}[t]
 \begin{center}
  \begin{tabular}{c|cc|cc}
   \hline
   cube & \# high-q & \% high-q & \# all & \% all \\
   \hline
   111 & 846 & 23.3\% & 985 & 21.4\% \\
   222 & 2171 & 59.9\% & 2743 & 59.7\%\\
   121 & 48 & 1.3\% & 64 & 1.4\%\\
   212 & 49 & 1.3\% & 74 & 1.6\%\\
   211 & 168 & 4.6\% & 255 & 5.5\%\\
   122 &  139 &  3.8\% & 216 & 4.7\%\\
   221 & 144 & 4.0\% &  169 & 3.7\%\\
   112 &  65 & 1.8\% & 94 & 2.0\% \\
   \hline
   TOT & 3630 & 100\% & 4600 & 100\%\\
   \hline 
  \end{tabular}
 \end{center}
\caption{Complete classification cube. The column ``cube'' contains the 3-digit identifier for the classifications adopted in this paper: first, second and third digit represent respectively spectral, photometric and morphological classifications.}\label{tab:cube}
\end{table}

This agreement confirms the goodness of this kind of classification: the vast majority of the galaxies in the sample belong to one of the two larger classes that show concordant behaviour in spectral, photometric and morphological properties. In these three fundamental observational features, bimodality is a major property of the galaxy population, both considering these features one at a time and comparing them in a more organic way.

\section{PCA-Clustering classification method}\label{sec:PCAclust}

The bimodality is an intrinsic property of galaxies, not only considering single specific characteristics like colours, spectral indices, morphologies etc, but also taking those properties as a whole, as we have seen in the previous section. A classification cube stands on its own because of this global bimodality, which tells us that galaxies are well divided in two categories, ``early types'' and ``late types''. How these two categories relate to each other is still matter of debate, and the characterisation of transitional galaxies -- objects that represent the bridge from one category to another, the so-called \emph{green valley} -- is of paramount importance for the definition of the evolutive history of the galaxies and to understand how and why galaxies migrate between categories.

For these reasons we decided to pursue a more global look to our sample, considering properties of galaxies as a whole. To accomplish this task, we used the Principal Component Analysis on our sample and a Cluster Analysis to identify the loci of \emph{early type} and \emph{late type} galaxies. 

\subsection{Principal Component Analysis}\label{subsec:PCA}

The Principal Component Analysis (PCA) \citep{Pea01, Hot33} is an orthogonal linear transformation useful to reduce multidimensional data sets to lower dimensions, in order to facilitate subsequent analysis. It transforms the data to a new coordinate system such that the greatest variance by any projection of the data comes to lie on the first coordinate (called the first principal component), the second greatest variance on the second coordinate, and so on.
For this reason PCA is the ideal tool to study a large number of parameters, allowing us to understand their importance and correlations.

\begin{figure}[t]
\begin{center}
  \includegraphics[width=.4\textwidth]{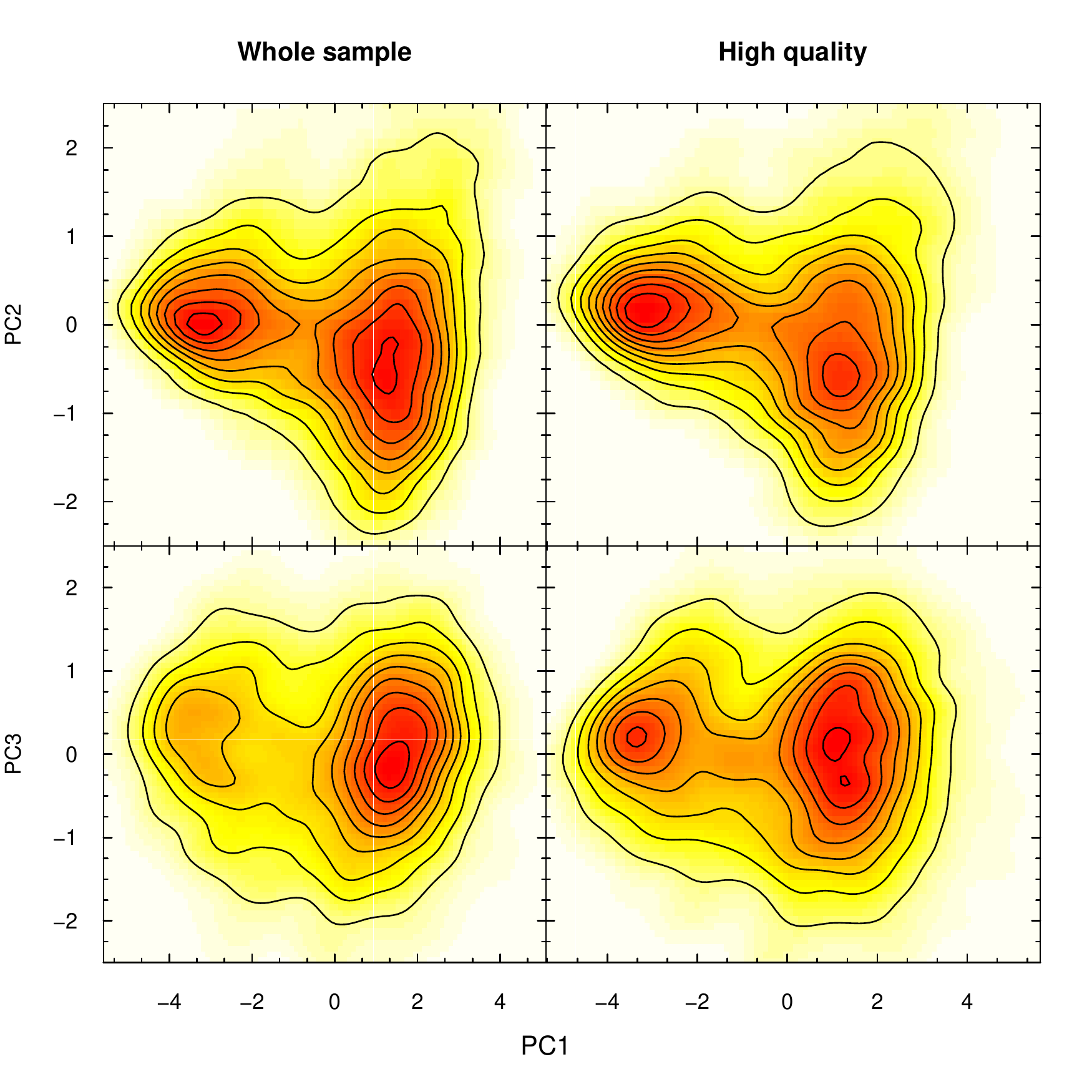}
\end{center}
\caption{2D density maps of the high redshift galaxies in PC1-PC2 plane (upper panels) and in PC1-PC3 plane (lower panels). Left maps are derived from the whole sample, while right ones are derived from the high quality sample only. It is clearly visible the global bimodality of galaxy properties, represented by the two ``clumps'' in density.}\label{fig:kernsmooth}
\end{figure} 

Our PCA run involved 8 major observational properties of the sample: two parameters are derived from spectra (the $D4000$ break and the $EW_0[\ion{O}{ii}]$); one is derived from the photometric analysis ($\Delta (B-\mathrm{z})$) and the remaining parameters are morphological: $M_{20}$ (second-order moment of the brightest 20\% of galaxy flux), concentration $C$ (ratio between radii including 80\% and 20\% of galaxy light), Gini coefficient $G$ (uniformity of light distribution), asymmetry $A$ (rotational symmetry of light distribution) and clumpiness $S$, as taken from ZEST catalogue. We chose these parameters in order to keep our results comparable to the previous classification, the 3D cube, which makes use of the same observables.

The first step required to apply the PCA to a data set is to normalise the involved observables. Thus, we took the logarithm of $EW_0[\ion{O}{ii}]$, as this variable is distributed as a log-normal distribution. Therefore, from now on we will be referring to $\log (EW_0[\ion{O}{ii}])$ every time we mention the equivalent width of~[\ion{O}{ii}].

The result of the PCA application to our eight variables is a rotated eight-dimensional space, where every new variable (PC$x$, where $x \in \mathbb{N}$, $x \leq 8$) is a linear combination of the original ones:
\begin{equation}
 PCx = \sum_{i=1}^8 a(i)_{x} V_i
\end{equation}

where $-1 \leq a(i)_x \leq 1$ are the coefficients of the linear transformation and $V_i$ are the original variables.

\begin{table*}[tb]
\begin{center}
 \begin{tabular}{lcccccccc}
  \hline
  Parameter&  PC1&    PC2&    PC3&    PC4&    PC5&    PC6&    PC7&    PC8 \\
   \hline
$D4000$    	& -0.368 &  0.117 &  0.423 &  0.062 & -0.653 &  0.329 & -0.365 & -0.026\\
$EW_{0}[\ion{O}{ii}]$ &  0.359 & -0.056 & -0.429 & -0.245 & -0.733 & -0.177 &  0.233 & -0.025\\
$\Delta (B-\mathrm{z})$ 	& -0.392 &  0.139 &  0.388 &  0.023 & -0.114 & -0.525 &  0.621 &  0.039\\
$G$     	& -0.367 &  0.304 & -0.415 &  0.031 &  0.002 & -0.571 & -0.522 & -0.038\\
$M_{20}$	&  0.419 & -0.013 &  0.323 &  0.131 & -0.058 & -0.314 & -0.261 &  0.730\\
$C$	 	&  0.400 &  0.125 & -0.289 & -0.447 &  0.065 &  0.320 &  0.160 &  0.640\\
$A$ 	   	&  0.185 &  0.772 & -0.160 &  0.488 & -0.028 &  0.234 &  0.215 &  0.066\\
$S$     	&  0.278 &  0.510 &  0.318 & -0.693 &  0.124 & -0.052 & -0.119 & -0.222\\
\hline
Prop. Variance  & 0.586 & 0.142 & 0.109 & 0.063 & 0.043 & 0.024 & 0.022 & 0.011\\
Cum. Variance & 0.586 & 0.728 & 0.838 & 0.901 & 0.944 & 0.968 & 0.990 & 1.000\\
\hline
 \end{tabular}
\end{center}
\caption{Results of the Principal Component Analysis applied to eight different properties of the galaxies. Absolute values of the coefficients show the relative importance of the original variables within each Principal Component; a negative coefficient means an anti-correlation. }\label{tab:PCA}
\end{table*}


In Table~\ref{tab:PCA} the coefficients $a(i)_x$ of our PCA are shown. Coefficients show the relative importance of the original variables in each eigenvector PC$x$: the larger the value of $a(i)_x$, the stronger the importance of the associated variable within the principal component. 
The two last rows of PCA table show the proportional variance (how much variance is expressed by each single PC) and the cumulative variance (how much variance is explained by \emph{the sum} of the previous PCs). We decided to never let the cumulative variance be below 80\% of the original total one, so we decided to keep the three first PCs, which explain 84\% of the original variance.

\begin{figure}[t]
\begin{center}
  \includegraphics[width=.4\textwidth]{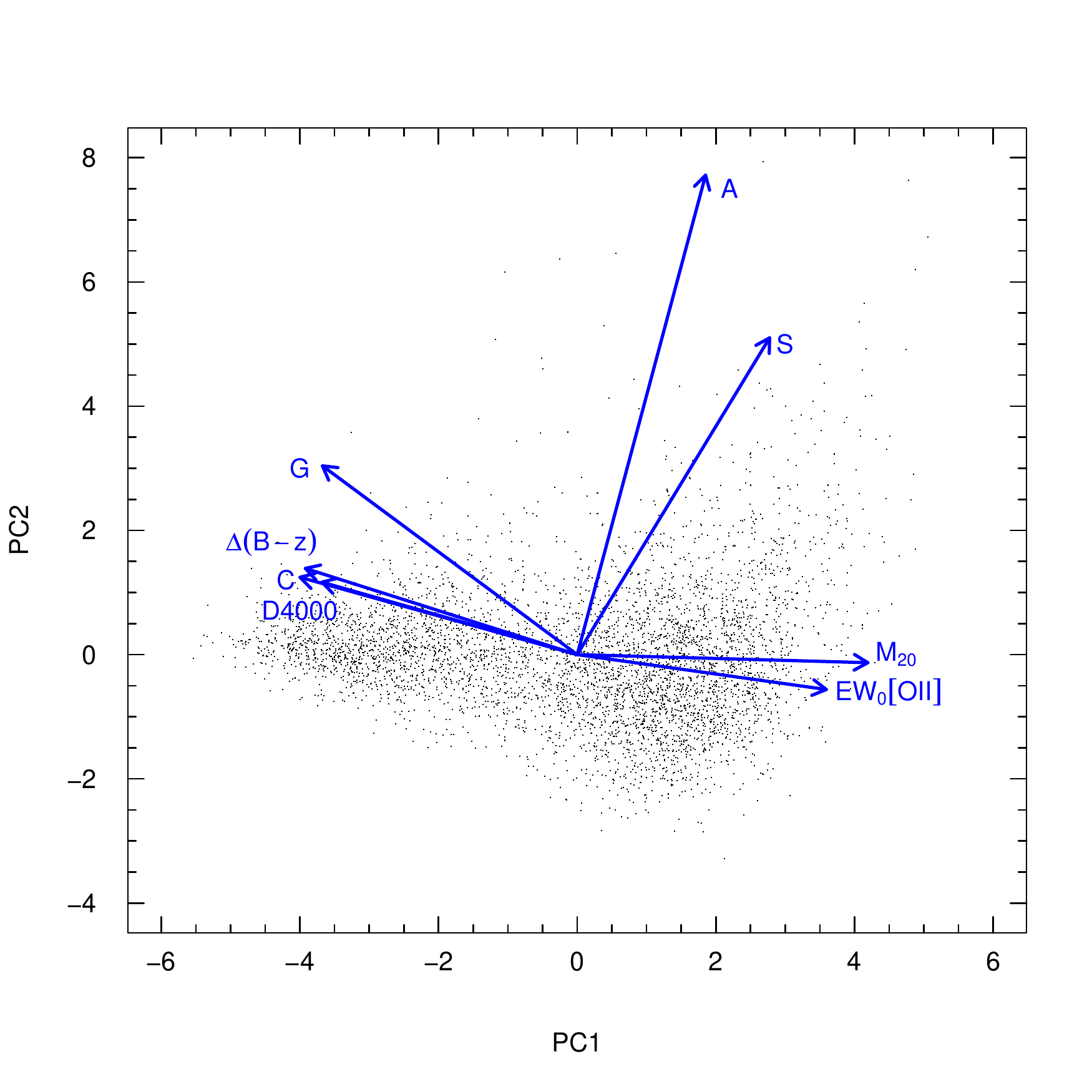}
\end{center}
\caption{Biplot of our PC1-PC2 plane. Black points are the galaxies as expressed in terms of PCs, while blue arrows represent the ``direction'' in which each original variable tends to scatter the data.}\label{fig:biplot}
\end{figure} 

Fig.~\ref{fig:kernsmooth} shows the density of the data points in the PC1-PC2 and PC1-PC3 planes, obtained via kernel density estimation with an axis-aligned bivariate normal kernel, evaluated on a square grid \citep{R:Venables+Ripley:2002}. The plot shows the isodenses of the points, both using lines of equal density and a colour-coded 2D map: the global bimodal nature of the whole population of galaxies is reflected by the two ``clumps'' in density, separated by a narrow under-dense ``valley'', in which transitional objects lie. The global bimodality is much more evident in the high quality sample, due to better measurements of the spectral features involved.

It is interesting to notice that \citet{Disney} stated that only one parameter should be sufficient to describe the nature of a galaxy, although they were not able to identify it: our PCA shows that the bimodality unfolds itself in the PC1 direction alone. Although PC1 cannot be that single simple parameter, it is a very interesting fact that the main properties of a galaxy can be described just by looking to its PC1 value.

The so-called biplot is a very useful tool to understand the relationships between the original variables and the PCs \citep{Gabriel}, and in our work it can help explain why do galaxies arrange themselves in this way in the PC space. In the biplot in Fig.~\ref{fig:biplot} the arrows represent the axes where each original variable lies, and their length is an index of their ``strength'', their importance within each PC -- in mathematical terms the coefficients $a(i)_x$ shown in Tab.~\ref{tab:PCA}, also called loadings. Looking at the coefficients of $D4000$, $EW_0[\ion{O}{ii}]$, $\Delta (B-\mathrm{z})$, $G$, $M_{20}$ and $C$ within PC1, for instance, one can see that they are roughly the same (in absolute value): this explains why in the biplot the relative arrows have more or less the same length along PC1 axis. 

Fig.~\ref{fig:biplot} shows that $D4000$ and $\Delta (B-\mathrm{z})$ are strongly correlated, because the arrows point in the same direction and have similar strength. The $EW_0[\ion{O}{ii}]$ is anti-correlated to both of them, and this is somewhat expected given the spectral classification shown in Fig.~\ref{fig:specclass}: most galaxies with high values of $D4000$ have little or no emission lines, and vice-versa. $\Delta (B-\mathrm{z})$ increases with $D4000$, so basically redder galaxies have a larger $D4000$, and this is also expected from Fig.~\ref{fig:photo}. We note also that $C$ and $G$ are strongly correlated: $G$ is a measure of how uniformly the flux is distributed among pixels in the galaxy image, so more concentrated galaxies have a larger value of $G$. $M_{20}$ is anti-correlated with the two other morphological parameters: since $M_{20}$ is a measure of how many bright off-centred knots of light are present, the greater is the value of $M_{20}$, the ``later'' is the galaxy, because disk-dominated galaxies have more bright spots (star formation regions, spiral arms, bars) than spheroidal or elliptical galaxies.

Taking into consideration only PC2 we can see that asymmetry $A$ and clumpiness $S$ are very strongly correlated: the larger the value of PC2 of a galaxy, the more disturbed its morphology is. Objects with low values of PC2 show more regular morphologies, and are separated by their values of the other morphological parameters like $C$, $M_{20}$ and $G$.

\subsection{Cluster analysis}\label{subsec:clusteranalysis}

Cluster analysis is based on partitioning a collection of data points into a number of subgroups, where the objects inside a cluster show a certain degree of closeness or similarity. Hard clustering assigns each data point (feature vector) to one and only one of the clusters, with a degree of membership equal to one, assuming well defined boundaries between the clusters. This model often does not reflect the description of real data, where boundaries between subgroups might be fuzzy, and where a more nuanced description of object's affinity to the specific cluster is required. For this reason we applied a fuzzy clustering method to our PCA-reduced sample in order to segregate galaxies between the two clusters. 

Our method makes use of the Unsupervised Fuzzy Partition (UFP) clustering algorithm as introduced and developed by \citet{Gath-Geva}. The approach of this method is Bayesian: first it is required to run a partition algorithm to provide first guesses of memberships and cluster centroids. This is achieved via a modification of the fuzzy K-means algorithm \citep{Bezdek73}. These prototypes are then used by the second algorithm (Fuzzy modification of maximum likelihood estimation -- FMLE) to achieve optimal fuzzy partition \citep{Geva2000511}.

\begin{figure}[t]
\begin{center}
  \includegraphics[width=.4\textwidth]{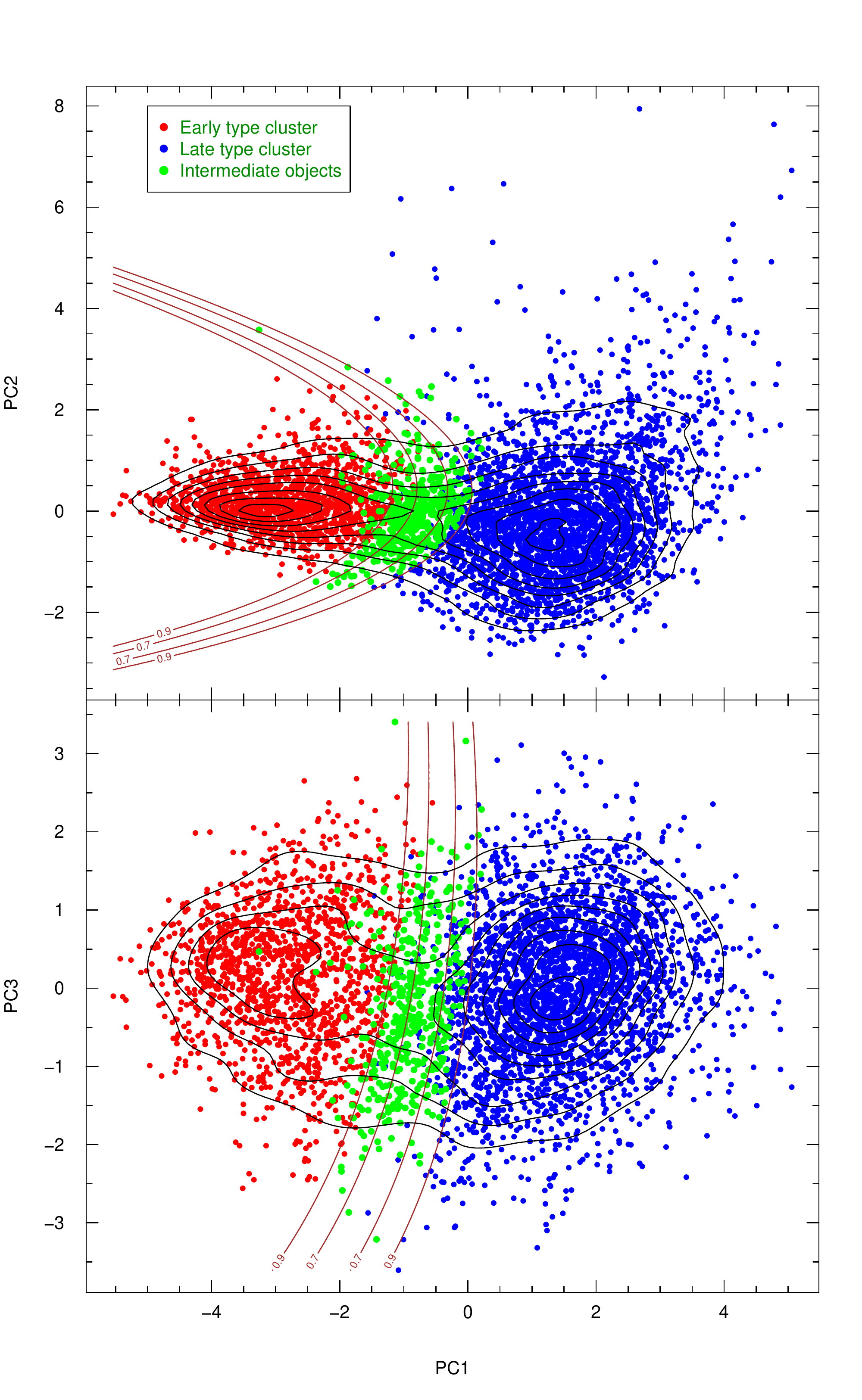}
\end{center}
\caption{Result of the Unsupervised Fuzzy Partition (UFP) clustering algorithm applied to the PCA-reduced whole sample: the upper panel represent the PC1-PC2 plane, while the lower panel represent the PC1-PC3 plane. In red are early type galaxies, in blue late type galaxies, in green our intermediate objects. Brown lines are the interceptions on both planes of the isoprobability surfaces with probabilities 70\% and 90\%. Black curves are the isodenses of the points in the planes, computed via gaussian kernel smoothing.}\label{fig:GG_PCA}
\end{figure} 

\begin{figure}[t]
\begin{center}
\subfigure{\includegraphics[width=.35\textwidth]{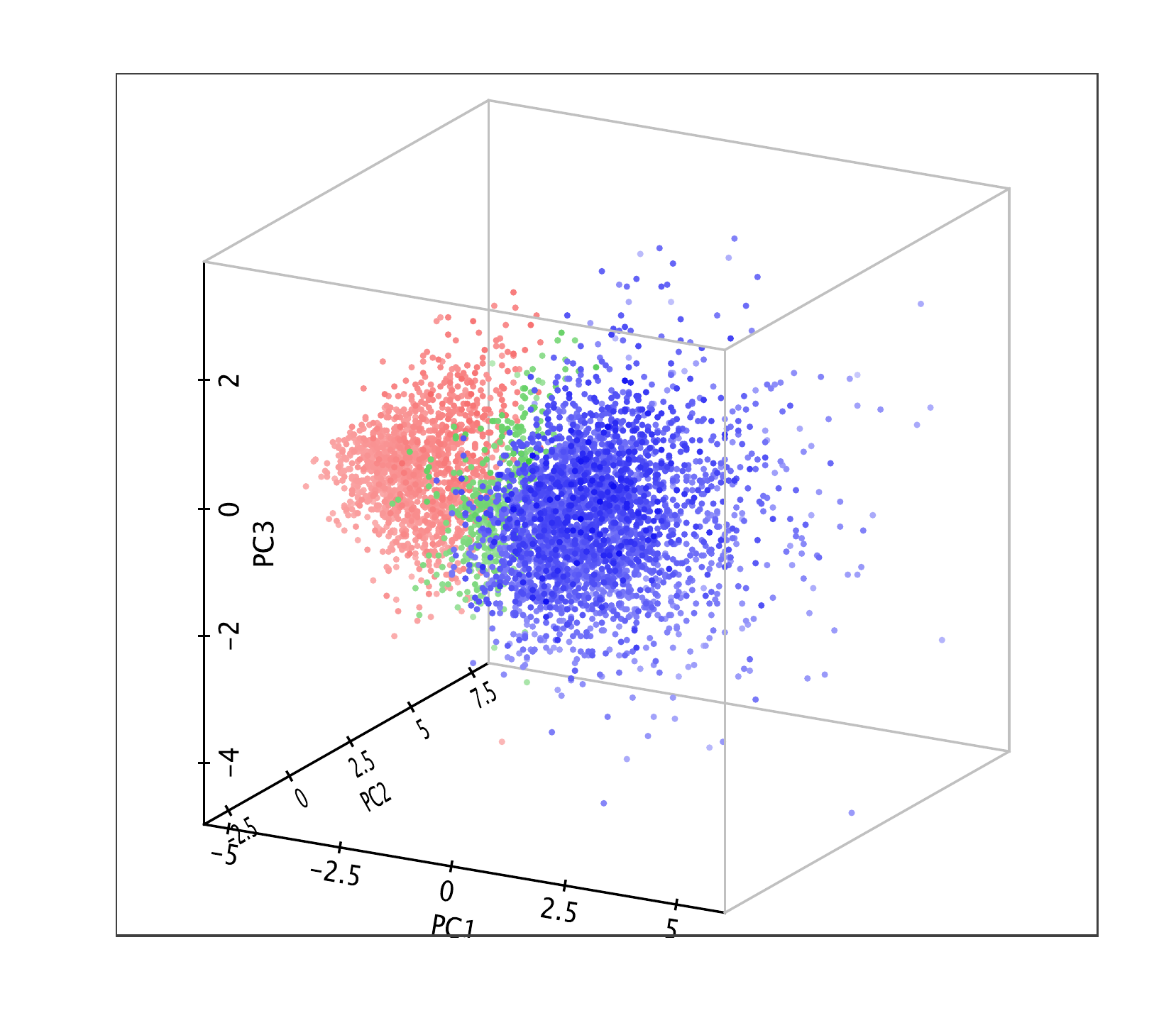}}
\subfigure{\includegraphics[width=.35\textwidth]{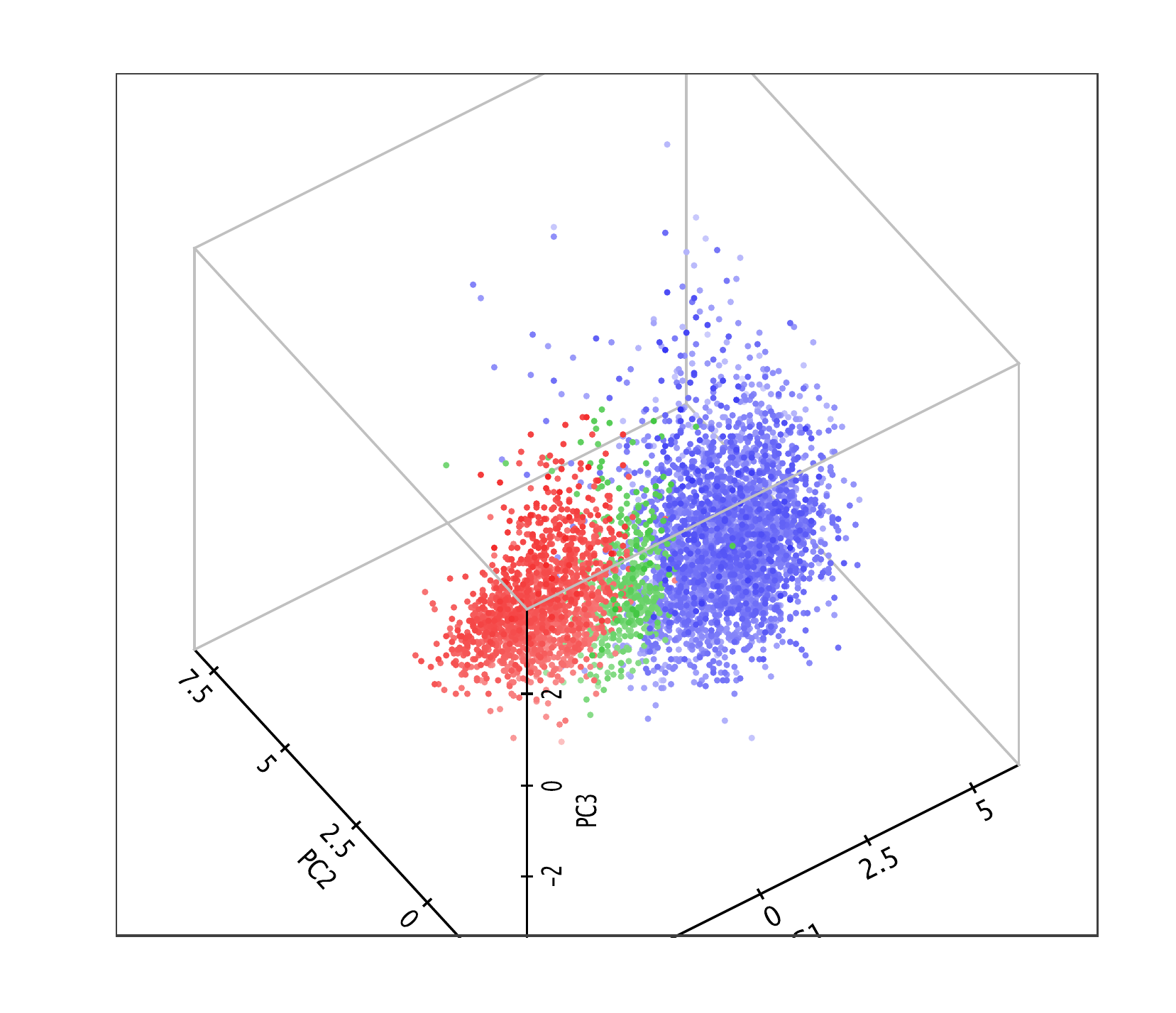}}
\end{center}
\caption{Two different three-dimensional visualizations of the PC space. The colours represent the clusters as defined by the UFP cluster analysis in Fig.~\ref{fig:GG_PCA}. Different intensities of the colours represent the distance of the point from the vantage point, trying to give the idea of the depth of the points distribution.}\label{fig:3D}
\end{figure}

Fig.~\ref{fig:GG_PCA} shows 2D projections of the application of the UFP clustering algorithm to our 3D dataset. The global bimodality shown by the PCA application is confirmed and well defined by the UFP algorithm. As already noticed in \S\ref{subsec:PCA}, the leftmost objects (in red in the plot) are the early type galaxies, while in the rightmost part of the diagram (in blue) are the late type galaxies. Figs.~\ref{fig:3D} are 3D visualization of the data, trying to show the PC-spatial distribution of the different galaxy populations.

Being a fuzzy partitioning method, objects do not belong just to one cluster: for any given data point, its probability of membership is spread across all the clusters, provided that the sum of memberships for all clusters is equal to 1. 

In our work we assign objects to a cluster only if their probability of membership for one of the clusters is $P>0.9$. We chose this threshold because, due to the exponential nature of the FMLE distance function, there is a steep rise in the probability function until $P \sim 0.9$, and then there is a general flattening for $P \gtrsim 0.9$. In Fig.~\ref{fig:GG_PCA} red objects are galaxies which belong to the ``early type cluster'' with a probability of more than 90\%, while blue objects are galaxies which belong to the ``late type cluster'' with the same probability threshold. All other galaxies (those which belong to a cluster with a probability $0.5<P<0.9$) are marked in green.

Early type galaxies, defined in this way, represent almost 30\% of the entire sample (1413 objects), while late types are 62\% (3035) and the other 8\% (426) are classified as intermediate objects. The early types' locus here is more populated than the correspondent class in the classification cube (the ''111'' class), which was composed by 23\% of the total sample (Table~\ref{tab:cube}). This is due to several reasons: the 90\% membership threshold for the UFP cluster analysis, which seemed a fair choice due to the shape of the probability function, is however more or less arbitrary; choosing a 95\% membership threshold, for instance, lowers the percentage of early type objects to $\sim 20\%$. Moreover, the classification cube considers 8 different classes of objects, while PCA+UFP only 3 of them: many of the outliers in the classification cube (all the 121s and the 211s, and a great part of 112s and 221s) are now classified as early types in PCA+UFP. If they were to be classified as fully concordant 111s in classification cube, this class would be made up of $ \sim 31\%$ of the whole sample. Finally, one must keep in mind that the ``early type cluster'', as defined by PCA+UFP, is not intended to be made up of pure passive galaxies; rather, it is composed also by bulge-dominated weakly-starforming objects.

Most of the differences between the two methods can be ascribed to errors and misclassifications due to the ``hard partitioning'' logic of the old cube classification: each of the sub-classifications of the cube were characterized by clear-cut boundaries that can produce placement misclassifcations, especially for objects that are in proximity of those boundaries. Another culprit could be the high number of morphological parameters in the PCA+UFP analysis, that might assign greater importance to those to the detriment of other parameters; however, several runs of the PCA+UFP algorithms with lower numbers of morphological parameters do not seem to substantially change the results.

Fig.~\ref{fig:GG_PCA} shows also the local density evaluation as shown in Fig.~\ref{fig:kernsmooth}. It can be easily seen that the intermediate objects lie in the ``valley'' between the two major clumps of data points. This is something expected, since we wanted to point out the relative difference between these objects and the galaxies belonging to the two clusters.

\subsection{Extension to low redshifts}\label{subsec:lowredshifts}

Due to the parameter choice of this analysis, we were forced to limit the analysis to a sub-sample of the 10k~zCOSMOS sample: as we said in \S\ref{sec:zcosmos}, the spectral features involved in the analysis ($D4000$ and $EW[\ion{O}{ii}]$) are detectable within zCOSMOS-bright only at $0.48 < z < 1.28$. The higher limit in redshift coincides with the limit of the zCOSMOS-bright survey, but the nearest galaxies (between $0 < z < 0.48$) were left out of the analysis. In order to expand the analysis and to follow the behaviour of galaxies in the entire redshift range of zCOSMOS-bright survey, we decided to exploit the PCA+UFP method to probe the galaxies even at lower redshifts, substituting the spectral features used at high redshifts with one of the best star formation indicators -- H$\alpha$ -- which is detectable within zCOSMOS-bright from the local universe to $z \sim 0.48$. This is one of the main reasons behind this work: the PCA+UFP method, not being tied to a particular set of data, is able to use different parameters and probe different redshift ranges and properties of the galaxies.

For the extension at low redshifts we therefore considered 7~observable parameters: $\Delta (B-\mathrm{z})$, $M_{20}$, concentration~$C$, Gini coefficient~$G$, asymmetry~$A$, clumpiness~$S$ and $EW_0(\mathrm{H}\alpha)$. Like in the previous analysis with $EW_0[\ion{O}{ii}]$ we considered the logarithm of the equivalent width due to its log-normal distribution, so from now on $EW_0(\mathrm{H}\alpha)$ has to be intended as $\log EW_0(\mathrm{H}\alpha)$. The \emph{low redshift sample} defined in this way is composed by 3402 galaxies. Results of the application of the PCA are shown in Table~\ref{tab:PCA_low}. As for the analysis at high redshifts, we decided to consider those PCs that give a cumulative variance not less than 80\%. In this case we took into account the first 4 PCs, which account for 89\% of the total original variance.

\begin{table*}[tb]
\begin{center}
 \begin{tabular}{lccccccc}
  \hline
  Parameter&  PC1&    PC2&    PC3&    PC4&    PC5&    PC6&    PC7\\
   \hline
$EW_0(\mathrm{H}\alpha)$& 	0.340& -0.097& -0.545& -0.541& -0.529& 0.032& -0.042\\
$\Delta (B-\mathrm{z})$&-0.404&  0.260&  0.311&  0.153& -0.766& 0.230& -0.085\\
$G$&   	       -0.439&  0.024& -0.463& -0.024&  0.249& 0.717&  0.119\\
$M_{20}$&       0.500&  0.060&  0.104&  0.220&  0.060& 0.471& -0.678\\
$C$&						0.216&  0.634&  0.358& -0.520&  0.178& 0.217&  0.269\\
$A$&	       	-0.471&  0.167& -0.045& -0.427&  0.186&-0.293& -0.666\\
$S$&  					0.086&  0.698& -0.499&  0.423& -0.007&-0.274& -0.004\\
\hline
Prop. Variance  & 0.483 & 0.177 & 0.126 & 0.104 & 0.060 & 0.035 & 0.014\\
Cum. Varariance & 0.483 & 0.660 & 0.786 & 0.891 & 0.950 & 0.986 & 1.000\\
\hline
 \end{tabular}
\end{center}
\caption{Results of the Principal Component Analysis applied to the low redshift ($z< 0.48$) galaxies.}\label{tab:PCA_low}
\end{table*}

\begin{figure}[t]
\begin{center}
  \includegraphics[width=.4\textwidth]{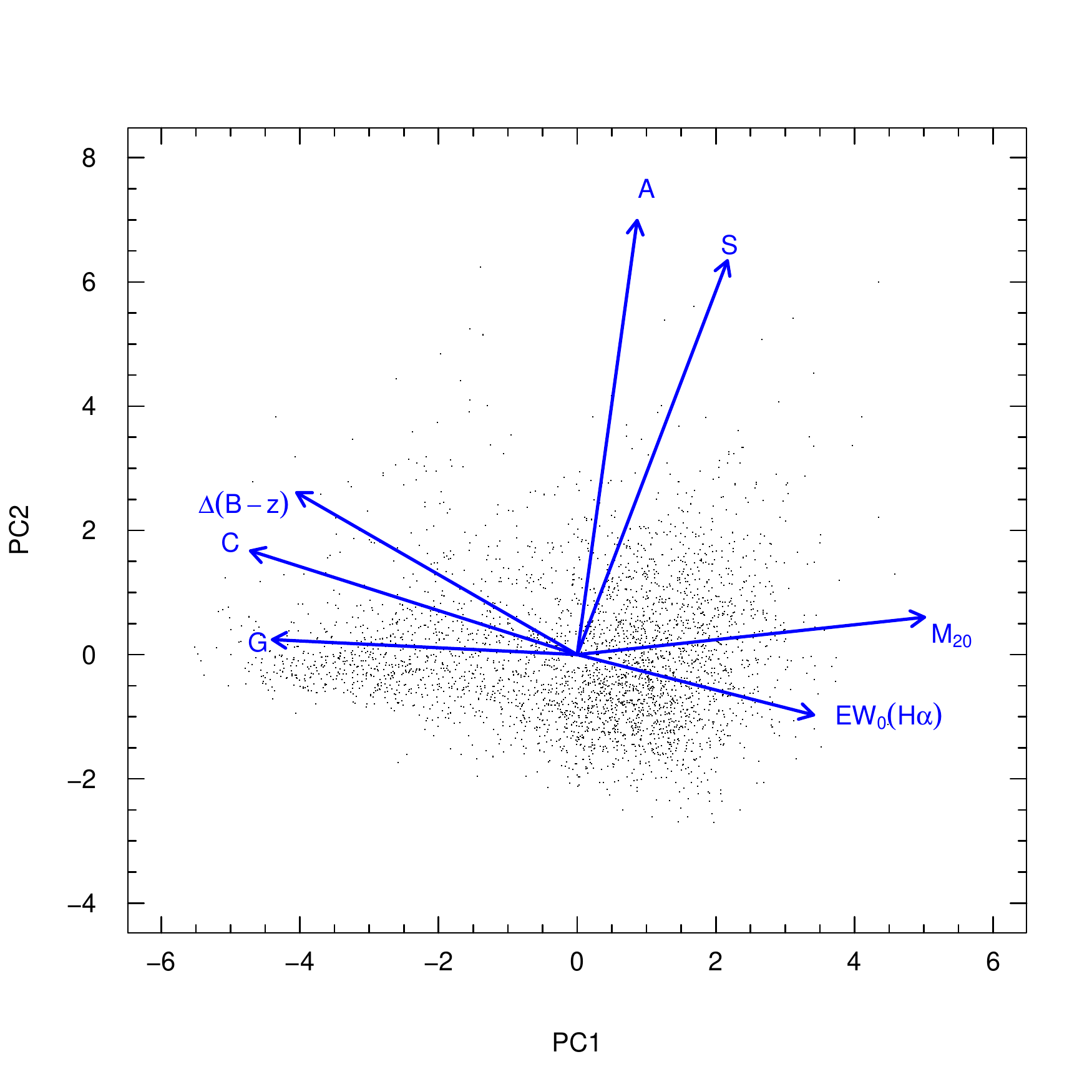}
\end{center}
\caption{Biplot of PC1-PC2 plane for low redshift galaxies.}\label{fig:biplot_low}
\end{figure} 

\begin{figure}[t]
\begin{center}
  \includegraphics[width=.5\textwidth]{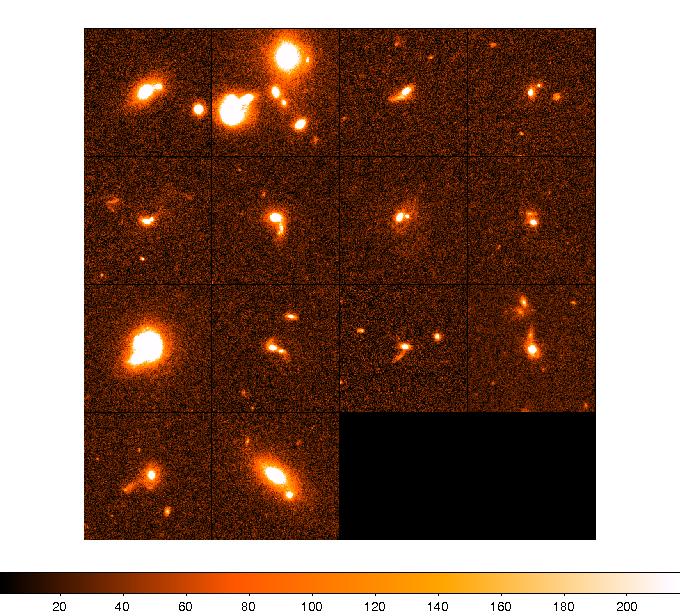}
\end{center}
\caption{Composite ACS image \citep[see][]{Koekemoer07} of low redshift early type galaxies with highest values of PC2. Their morphologies are quite complex, suggesting tidal interactions and recent merging.}\label{fig:PC2_magg2}
\end{figure}

In Fig.~\ref{fig:biplot_low} the biplot of the PCA for low redshift galaxies is shown. By comparing it with Fig.~\ref{fig:biplot} one can see the striking resemblance in the cloud's shape and in loadings' directions. The function of $D4000$ and $EW_0[\ion{O}{ii}]$ -- to segregate the galaxies mainly in PC1 direction -- is taken over by $EW_0(\mathrm{H}\alpha)$, while the other parameters' relations remain largely unchanged. With respect to Fig.~\ref{fig:biplot}, galaxies in the early-type cluster spread more in PC2 (which is mainly morphology driven): this is probably due to ACS being progressively abler to recognise features, even in spheroidal galaxies, with decreasing redshift, due to the larger size of the galaxies themselves. So spheroidal galaxies with streams due to encounters with companions, interacting galaxies or just objects with companions nearby, have larger values of asymmetry $A$ and clumpiness $S$ with respect to galaxies with similar features but at higher redshifts (angular dimensions of those galaxies will be smaller and their features will most likely be too small and faint to be appreciated with an automatic analysis). This is evident in Fig.~\ref{fig:PC2_magg2}), where ACS snapshots of the galaxies in early types' cluster with highes values of the second principal component ($\mathrm{PC2} > 2$) are shown.

\begin{figure}[t]
\begin{center}
  \includegraphics[width=.4\textwidth]{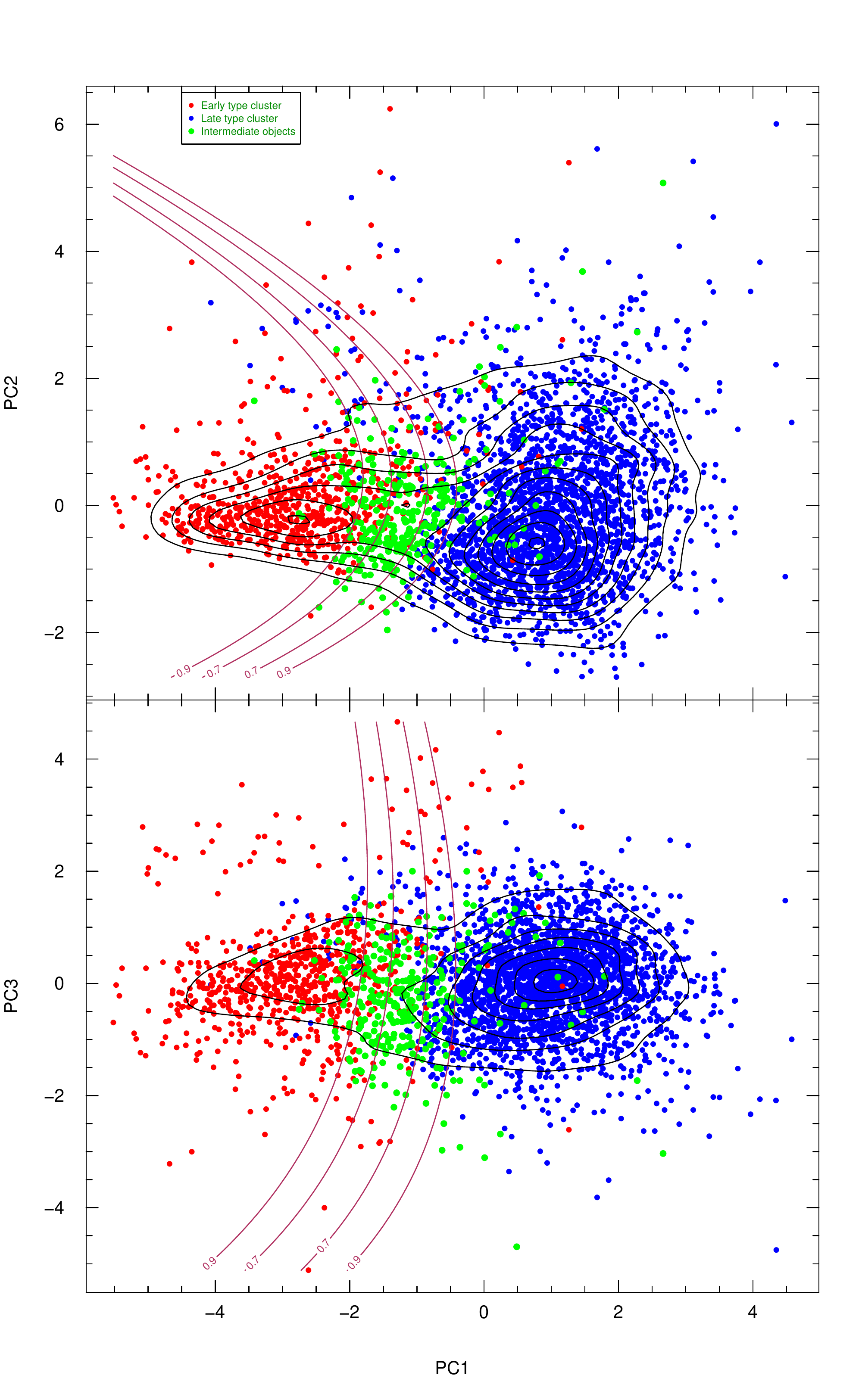}
\end{center}
\caption{Cluster analysis results for low redshift galaxies. Superimposed to the points, as in Fig.~\ref{fig:GG_PCA}, are the isodenses of the points calculated via kernel smoothing in PC1-PC2 and PC1-PC3 planes. The curved lines represent the projected isoprobability curves. Clusters and green valley objects appear more scattered across the planes because of projection issues from four-dimensional PCA to the 2 dimensions of the plot.}\label{fig:cluster_low}
\end{figure}

Fig.~\ref{fig:cluster_low} shows the result of the UFP clustering algorithm application to the low redshift sample of galaxies. As in previous analysis for the high redshift sample, we used a threshold of 90\% membership to distinguish between objects belonging to the ``early-type'' cluster, to the ``late-type'' one or objects not belonging to any cluster -- our ``green valley'' galaxies. Green valley objects lie in the saddle between the two main clusters, as it can be seen in the plot represented by isodenses, calculated by gaussian square kernel smoothing of the PC1-PC2 and PC1-PC3 planes, in a way similar to that of the high redshift galaxies (Fig.~\ref{fig:GG_PCA}). With respect to high redshift galaxies, clusters of low redshift galaxies appear less centred and defined: green dots, for instance, appear well beyond the boundaries of 90\% isoprobability that define them. This is due to the isoprobability curves being merely 2D projections of 4D hypersurfaces, since, as we said, we considered the first 4 PCs for the cluster analysis. 

Out of the 3402 objects the low redshift sample is made up of, early type galaxies represent 20.6\% (704 objects), while late type galaxies are 70.5\% (2401), and the green valley galaxies are 8.9\% (297). With respect to the high redshift sample, green valley objects represent more or less the same percentage of objects, while there is significant shift of populations between the two main clusters: late type galaxies are $\sim 10\%$ more with respect to the high redshift sample, while conversely early types are 10\% less. This is likely to be due to a selection effect (at low redshift we are sampling galaxies with lower luminosities and lower masses, which are on average ``later'' at all redshifts), rather than a real evolutive feature. In the next section we will explore in more details the evolution of the galaxy populations with redshift.

\section{Results}\label{sec:results}

The PCA+UFP analysis presented in this work offers many improvements with respect to the previous methods of classification like the classification cube. One of the greatest advantages of such an approach is given by its self-consistency and its global approach to the parameters: as we stated in \S\ref{subsec:clusteranalysis} the classification cube is prone to errors in one or more of its sub-classification methods because they are ``hard partition'' ones. Given the fact that every parameter is treated separately from the others, it is easier to have one of them misclassified due to internal errors or closeness of the value to the boundaries.

The PCA+UFP method reduced the possibility of this kind of errors because its parameters are treated simultaneously: using the PCA on a multidimensional space we are ``averaging out'' outlying values in a small number of parameters. This can be intuitively understood by looking at biplots (Figs.~\ref{fig:biplot} and \ref{fig:biplot_low}): an outlying value in $M_{20}$, for instance, can be compensated by ``normal'' values in spectral emission lines, $D4000$ and $C$.

Another powerful feature of the PCA+UFP analysis is its flexibility: while the classification cube is strongly bound to its defining parameters -- and for this reason has been applied to the high redshift sample in this work -- the PCA+UFP analysis is not restricted to a particular dataset or a particular set of parameters. We therefore can extend the work to low redshifts just by substituting the two spectral parameters with a different one. The choice of H$\alpha$ has been made in order to keep the possibility to compare the results of high and low redshift samples, and have a comprehensive look to the whole 10k dataset. Actually, the PCA+UFP method can successfully be applied also to completely different datasets (star formation rates, masses, luminosities) of this or other galaxy surveys, and that is possibly its most important achievement.

In the next subsections we will show some of the properties of the whole 10k population, and of few interesting subsamples, in PCA+UFP analysis.

\subsection{Combined high and low redshift sample}\label{subsec:whole}

\begin{figure*}[t]
\begin{center}
  \includegraphics[width=.9\textwidth]{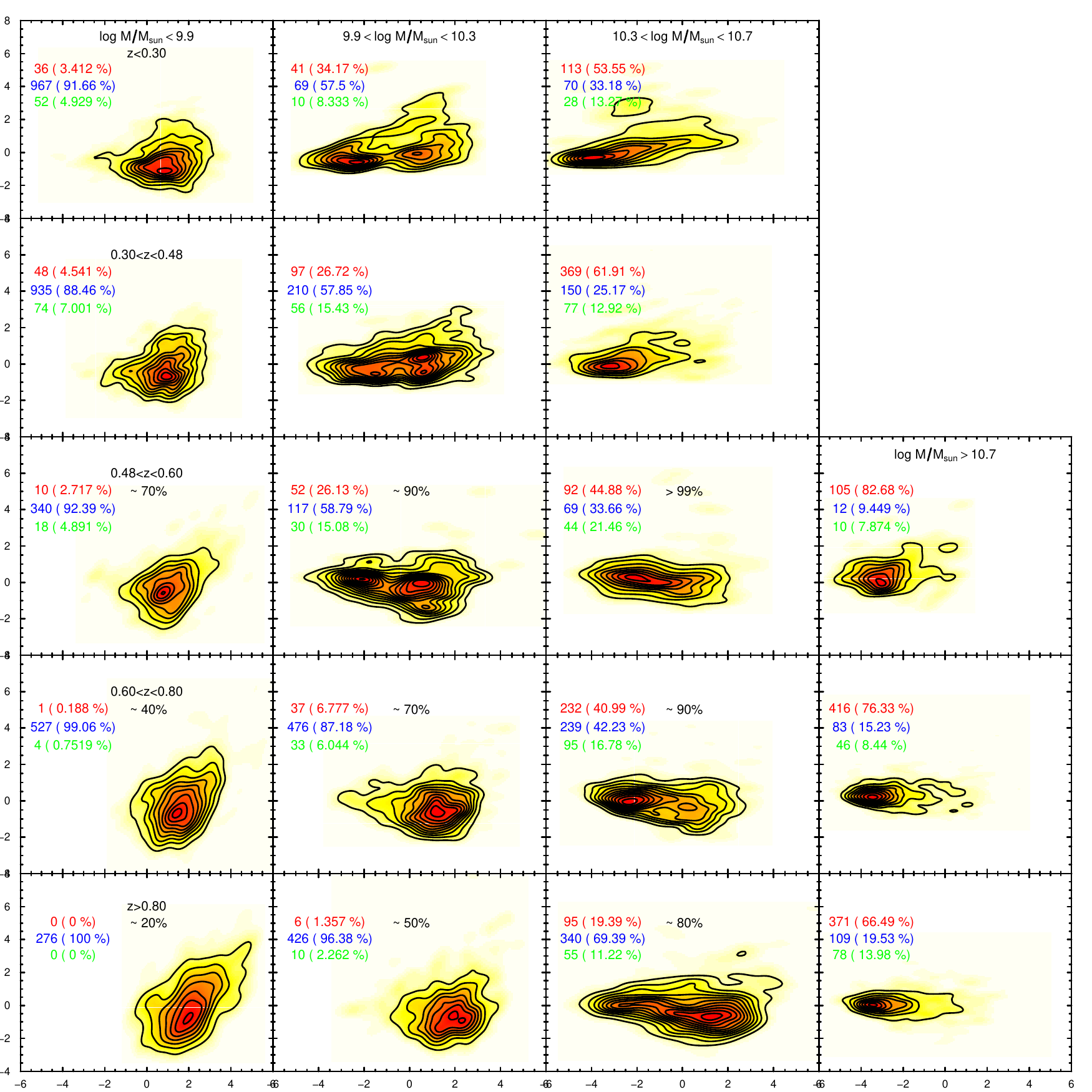}
\end{center}
\caption{PC1-PC2 diagrams for low redshift (upper two rows) and high redshift (lower three rows) samples, kernel smoothed with the usual technique. Columns represent bins of mass (growing from left to right, 
as specified inside first row boxes), while rows represent bins of redshift (growing from top to bottom, as specified in first column boxes). In each panel are also shown the absolute numbers and fractions of galaxies in each cluster (early-type, late-type and green valley), in red, blue and green respectively. In some of the high redshift panels are shown the mass completenesses \citep[as computed by][]{Pozzetti}; where there are no percentages the sample has to be intended as mass-complete.}\label{fig:enorme}
\end{figure*} 

Fig.~\ref{fig:enorme} shows the evolution of the different populations of galaxies, within the whole 10k sample, with redshift and with mass. Masses have been computed by \citet{Bolzonella09}, using \citet{BC03} population synthesis models, by means of the \verb|Hyperzmass| code, a modified version of the photo-$z$ code \verb|Hyperz| \citep{Bolzonella00}. 

Low mass galaxies ($\log M/\msun < 9.9$, first column) are almost exclusively part of the late-type cluster, while high mass galaxies ($\log M/\msun > 10.7$, last column) mainly belong to the early-type cluster. The transition can be mostly seen in the intermediate mass bins: at $9.9 < \log M/\msun < 10.3$, galaxies at high redshift ($z >0.80$) are still forming stars actively, and are therefore concentrated in the late-type cluster; the ``migration'' towards the early type cluster seems to begin at moderately lower redshifts ($0.60 < z < 0.80$), slowing down from $z \sim 0.50$ and being still ongoing also in the local Universe. 

\begin{figure*}[t]
\begin{center}
  \includegraphics[width=.7\textwidth]{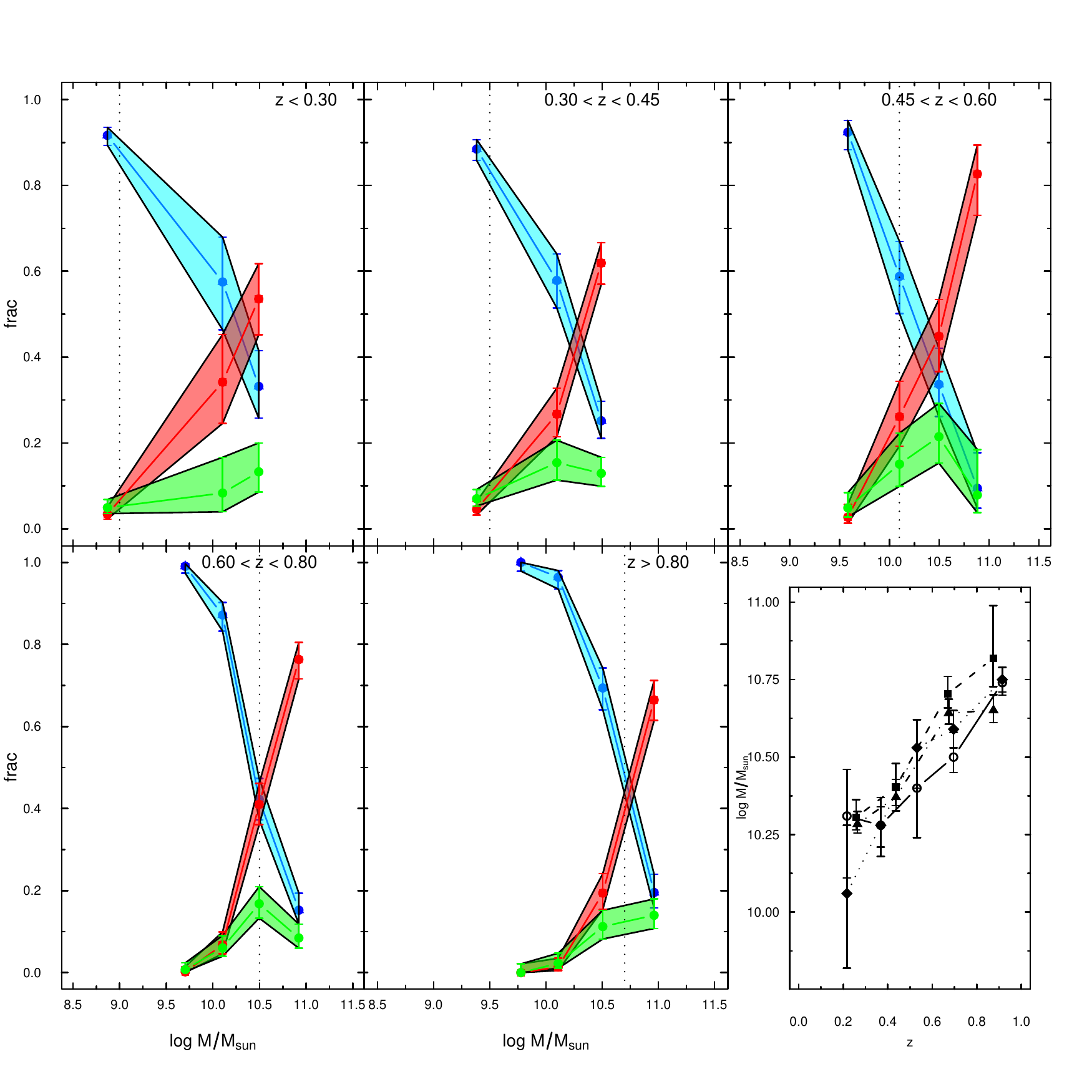}
\end{center}
\caption{Evolution with redshift of the fractions of different galaxy populations in mass. Each panel shows the fraction of galaxies in each mass bin that belong to each PCA+UFP cluster (in cyan are late-type galaxies, in red the early-type ones, in green the green valley ones), in a specific redshift bin. Errors are 95\% confidence intervals for multinomial populations \citep{nla.cat-vn712276}. Vertical dotted lines represent the 90\% mass completeness in each redshift bin. The last panel represents the evolution in $z$ of the transition mass ($M_{\mathrm{cross}}$), defined as the point where red line and cyan line meet (open circles and solid line). Errors associated are given by the width of the region where the two strips meet. Dashed 
and dot-dashed lines represent the transition masses as calculated in \citet{Pozzetti}, respectively using Marseille morphologies 
and SED colours photometric classifications. The dotted line represents the transition masses as calculated using \citet{Balogh} definition of green valley applied to our combined sample (see \S\ref{subsec:greenvalley}).}\label{fig:masstransition}
\end{figure*} 

At slightly larger masses ($10.3 < \log M/\msun < 10.7$) this transition appears to happen at earlier epochs: at $0.60 < z < 0.8$ early-type and late-type galaxies are numerically comparable, and the transition appears almost complete at $0.30 < z < 0.45$. At very low redshifts ($z < 0.30$) the percentage of late-type galaxies seems to rise again: this is most likely due to the effect of asymmetry $A$ and clumpiness $S$ in low-redshift ACS images we mentioned in \S\ref{subsec:lowredshifts}. This delay in the star formation quenching for the lower mass galaxies, in opposition to the larger ones, can be regarded as one manifestation of the \emph{downsizing} effect: the main reasons behind this effect are still unclear, even if some mechanisms have been suggested \citep{Bower, Hopkins, Dekel}. Some numerical simulations \citep{Schweizer} show that the transition in colours should be very fast (of the order of $\sim 500$ Myr), and other observational studies seem to suggest that this is the case if the star formation is quenched efficiently; \citet{Balogh}, however, showed that an exponentially decaying star formation can lengthen the transition phase to some Gyrs. Our work seem to suggest that a global transition (from our ``late type'' locus to the ``early type'' one) takes longer to be achieved (at least some Gyrs). Part of this is certainly due to the changes in colours and morphologies taking place with different timescales.


Looking at Fig.~\ref{fig:enorme} by rows it is possible to appreciate the mass distribution of the galaxy population at fixed redshifts. At low redshifts the zCOSMOS survey cannot sample the high mass galaxies ($\log M/\msun > 10.7$) due to the small sampled volume and the bright magnitude cut, so the corresponding boxes are empty. 
At higher redshifts mass incompleteness prevents us to directly compare the numbers of galaxies in each mass bin (as it can be seen in the plot, at $z>0.80$ the mass completeness of the sample with $\log M/\msun < 9.9$ is of the order of 20\%). However, this is not a severe issue when dealing with fractions within each mass and redshift bin; 
we can assume that within the bin the mass distribution is rather flat. However, due to mass incompleteness the highest redshift and lowest mass bins are to be considered with caution.

We summarise these considerations in Fig.~\ref{fig:masstransition}, 
where each of the first five panels represents a row of Fig.~\ref{fig:enorme}, i.e. a bin of redshift in which we divided our sample. For every given redshift bin the fraction of early type, late type and intermediate objects for each mass bin are plotted. Low mass early type galaxies are very few ($\sim 4\%$) in every redshift bin, late types being by far most frequent at $\log M/\msun < 9.9$, as it can be seen also in the first column of Fig.~\ref{fig:enorme}.  This is in good agreement with determinations of \citet{Kovac10} for the same zCOSMOS sample, who found a similar behaviour in different environments for galaxies of different morphological type.

Intermediate objects seem to be numerically important around $\log M/\msun \sim 10.5$ at high redshifts, constituting up to $\sim 20\%$ of the sample at $z \sim 0.5$. This suggests that the evolutive transition from the blue cloud towards the red sequence may be most important at intermediate redshifts and intermediate masses (central quadrants in Fig.~\ref{fig:enorme}).

From Fig.~\ref{fig:masstransition} the masses at which early-type and late-type galaxies are numerically the same at different redshifts ($M_{\mathrm{cross}}$), can also be derived. This transition mass $M_{\mathrm{cross}}$ is plotted in the lower right panel of Fig.~\ref{fig:masstransition} as a function of redshift. Transition masses computed in this work (solid line in the plot) are in fair agreement with those calculated by \citet{Pozzetti} using Marseille morphologies \citep{Cassata07, Cassata08, Tasca} as separators of different galaxy types -- dashed line in figure -- and using a photometric classification \citep{Zucca} -- dot-dashed line. A Cram\'er-von Mises test \citep{Anderson1962} confirms the consistency of the three estimates of $M_{\mathrm{cross}}$ (p-values above 0.73).
It must be kept in mind, though, that determinations of $M_{\mathrm{cross}}$ in this work are made within a three-cluster framework (early type, late type and intermediate galaxies), while other determinations are made taking into account only the two main galaxy populations. Splitting our intermediate galaxy sample between the other two clusters, using a 50\% threshold as membership values, the evolution with redshift of $M_{\mathrm{cross}}$ steepens, and especially at high redshifts transition masses are even more in agreement. Considering the different techniques of calculation, however, the agreement among these determinations is quite remarkable.

\subsection{Green valley galaxies}\label{subsec:greenvalley}

\emph{Green valley} galaxies have been defined in a number of different ways, usually exploiting their natural bimodal distribution using colour indicators like $u-r$ \citep{Strateva, Baldry04}, $U-V$ \citep{Brown, Silverman}, $U-B$ \citep{Vergani}, $B-i$ \citep{Caputi}. In this subsection we will analyse the $U-V$ rest-frame colour distribution (from now on $(U-V)_0$) of our PCA+UFP clustered galaxies.

\begin{figure}
\begin{center}
  \includegraphics[width=.5\textwidth]{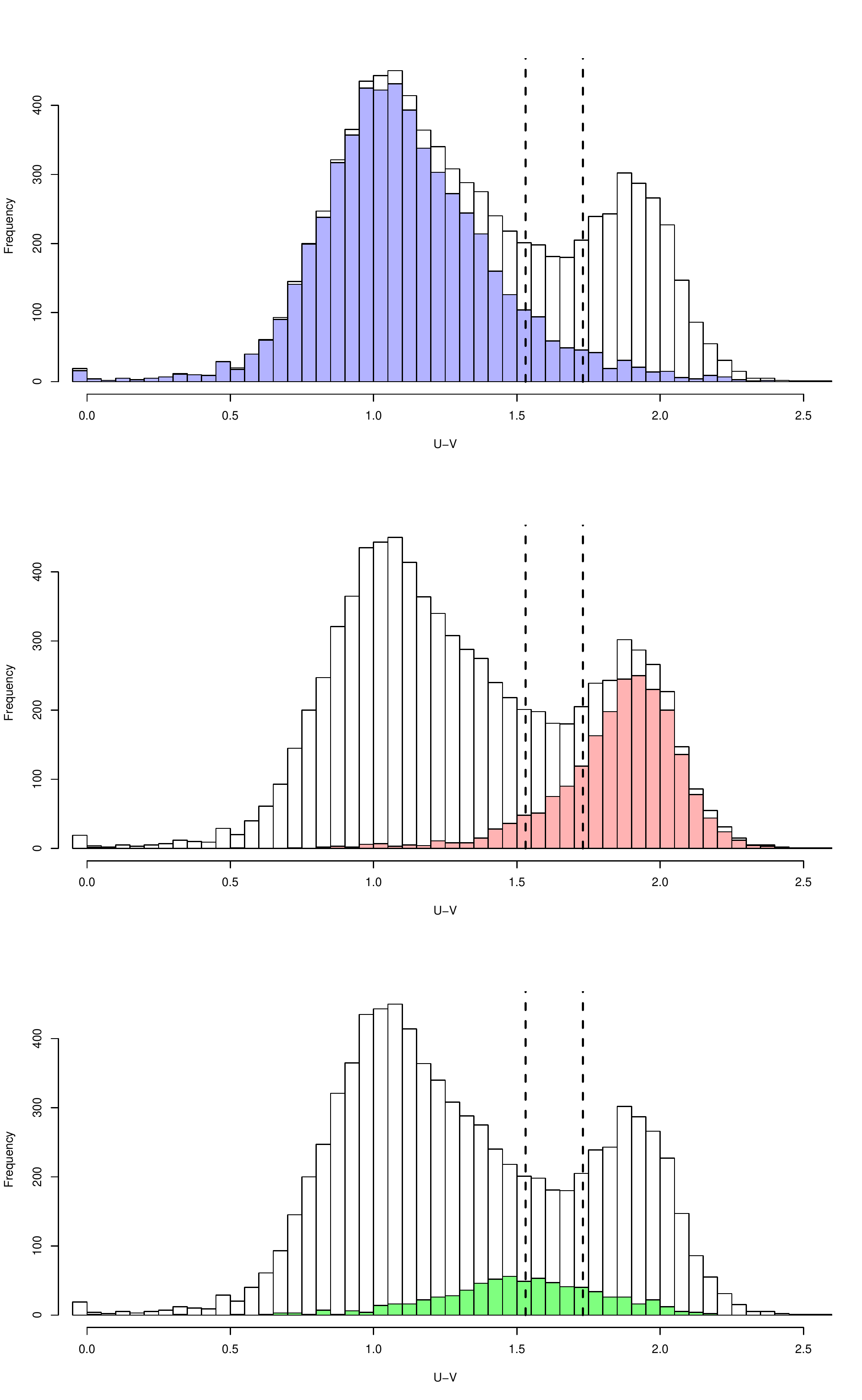}
\end{center}
\caption{Rest frame $U-V$ distributions of the galaxies in the combined sample (high+low redshift). Open histograms represent the distribution of the total sample; blue, red and green histograms represent the distribution of PCA+UPF late types, early types and intermediate galaxies, respectively. Dashed lines represent green valley boundaries as defined by \citet{Balogh} for comparative purposes.}\label{fig:UmenV}
\end{figure}

The $(U-V)_0$ distribution of the combined high+low redshift samples (Fig.~\ref{fig:UmenV}) shows a clear bimodality, that reflects the global one we discussed throughout the paper. The separation between the two families in colour happens at $(U-V)_0 \sim 1.6$; the colour distribution of our late type galaxies peaks at $(U-V)_0 \sim 1$, while the distribution of the early types is peaked at $(U-V)_0 \sim 1.9$. All of these are in fair agreement with other determinations from literature \citep{Silverman, Brammer}. The green valley objects' distribution is peaked at $(U-V)_0 \sim 1.5$, near the saddle of the total distribution.

We can compare the $(U-V)_0$ distribution of our green valley galaxies with \citet{Balogh}  definition of green valley, which is defined as the $0.2\; \mathrm{mag}$ dip between the two observed Gaussian distribution for early- and late-type galaxies.
Applying the above definition, in the combined sample 760 objects out of 8\,256 (9.2\%) would be defined as ``green valley'' objects; this number is very close to the number of green valley galaxies in our classification (721, 8.7\%); more than 25\% of our green valley objects are so also in the \citet{Balogh} definition, while the rest of the objects within those boundaries are almost equally divided by PCA+UFP between the two main clusters. The largest part of our intermediate galaxies lies to the left of the colour-defined green valley, i.e.\ in the region of the blue galaxies, but makes up only 6.5\% of all the objects in that region; conversely, PCA+UFP intermediate galaxies constitute 8.4\% of all the objects in the red galaxies region. 

Being based on overall properties of the galaxies, our classification method gives somewhat different results compared to classical colour definitions of green valley: the cores of the early-type and late-type clusters are correctly reproduced, but our classification suggests that relying on a single colour might not be sufficient to correctly recover those galaxies which are really in transition between the late-types and the early-types clusters.

The transition masses $M_{\mathrm{cross}}$ of the sample divided using \citeauthor{Balogh} definition of green valley were also calculated (dotted line in Fig.~\ref{fig:masstransition}); the agreement between the determinations is very high, even considering the uncertainties in the first redshift bin due to the low number of objects. Using a mass and/or redshift dependent colour definition of the green valley \citep[e.g.][]{Brand09} results are very similar.

\subsection{Red spirals}\label{subsec:redspirals}

We checked the PCA+UFP clustering properties of some of the outliers in the classification cube. Obviously this has been possible only with galaxies from the high redshift sample, because the classification cube has been defined using $D4000$ and $EW_{0}[\ion{O}{ii}]$, which were available only at $z>0.48$ (see \S\ref{subsec:spectral}). \emph{Red spirals}, for instance, are often identified with edge-on spiral galaxies, reddened by a strong dust lane \citep{Zucca, Tasca}, while face-on red spirals are thought to be the very oldest spirals which used up their gas reservoirs, probably aided by strangulation and bar instabilities \citep{Masters}. In our classification cube, red spirals may be identified by the three-digit codes ``112'' and ``212'', both representing morphological late-type galaxies (third digit ``2''), the first one representing spectrally passive red objects and the latter one referring to red star-forming galaxies.

Galaxies with classification cube code ``112'' are 93: of those, 24 (25.8\%) are classified by PCA+UFP in the green valley group; 27 (29\%) are in the late type cluster; 43 (46.2\%) are in the early type cluster. A fairly high number of them (14) possess unusually high values of PC2: at a visual inspection those objects revealed very disturbed morphologies, dominated by merging and tidal streams (Fig. \ref{fig:PC2_magg2}), in agreement with determinations from \citet{Conselice00} who found that very large values of $A$ (reflecting in our work in large values of PC2) are a good indication of ongoing major merging. At least for these objects, automatic morphological classification methods apparently fail to identify correctly them as merging spheroidals: their asymmetric characteristics are instead intepreted as late type morphologies.

Galaxies with classification cube code ``212'' are 74: 25 of them (33.8\%) are classified in the green valley group, 43 (58.1\%) are in the late type cluster and only 6 (8.1\%) are classified in the early type cluster. Their range in PC1 and PC2 is quite narrow, making those objects a rather homogeneous sample, located in the middle of the PC1-PC2 diagram, in or very near the low density saddle between the clusters. Those galaxies, showing spiral morphologies, low star formation rates (indicated by PC1 $\sim 0$) and reddish colors are the best candidates of the old spirals population mentioned by \citet{Masters}.



\subsection{Blue ellipticals}\label{subsec:blueellipticals}

In our classification cube, blue ellipticals are identified by the three-digit codes ``121'' and ``221'', the first one representing spectrally passive objects and the latter one referring to active star-forming galaxies, both bulge-dominated.

Classification cube code ``121'' galaxies are almost exclusively assigned to the early type galaxies cluster by the PCA+UFP algorithm (60/64), while code ``221'' show a somewhat diverse behaviour, being equally divided among the groups: 56 out of 169 (33.1\%) belong to the green valley group, 52 (30.8\%) to the late type cluster and 61 (36.1\%) to the early type cluster. In PCA terms, objects in the latter group are characterised by positive values of PC2 and generally negative values of PC1: while code ``121'' galaxies are most probably the result of a color misclassification in the classification cube, and therefore are ``normal'' early type galaxies --- confirmed by their $\Delta (B-\mathrm{z})$ very close to the dividing line in Fig.~\ref{fig:photo} --- code ``221'' objects seem to be more complex. Late type ``221''s have large values of PC2, while the PC2 value of early type ``221''s is around 0. This may imply a misclassification in $\Delta (B-\mathrm{z})$, too, but it is not sufficient to explain all their features. Most probably many of these objects, especially at higher values of PC1, present complex morphologies and are the result of tidal interactions.


 These results seem to imply that for these objects the spectrophotometric properties are given more importance than the morphological ones by PCA+UFP algorithm. In fact, as we said, a spiral morphology classifier -- especially when using wide classifiers and automatic recognition systems -- is more subject to errors due to the asymmetries of merging objects.



\subsection{Active Galactic Nuclei}\label{subsec:AGN}

We also investigated the positions, in the PCA spaces, of known AGN in the zCOSMOS sample. Type-1 AGN, which are easily recognisable by their broad emission lines, are given a particular spectroscopic confidence class since the determination of their redshifts 
and have been excluded from the subsamples. Type-2 AGN, on the other hand, are included in the sample since they are more difficult to identify, because their emission lines are very similar to those of regular star-forming galaxies. We used the diagnostic diagram selection of \citet{Bongiorno} to identify Seyfert 2 galaxies and LINERs and investigate their positions in PCA planes. Two different diagnostic diagrams have been exploited to select type-2 AGN, at low redshift using the line ratio [\ion{N}{ii}]/H$\alpha$ and [\ion{O}{iii}]/H$\beta$ whereas at high redhift the line ratios [\ion{O}{iii}]/H$\beta$ and [\ion{O}{ii}]/H$\beta$ have been used. Unfortunately, the different ionization properties of Seyfert~2 and LINERs galaxies are separable only using the diagnostic diagrams only at low redshifts. For this reason we will discuss the properties of the whole type-2 AGN population (which includes both active galaxy classes) in the two redshift ranges, separating the LINERs and Seyfert~2 galaxies only for $z\lesssim 0.5$ \citep[for a more detailed analysis see][]{Bongiorno}.

The analysed sample is composed by 79 type-2 AGN in the high redshift range and 125 type-2 AGN (95 of which are LINERs, while the other 30 are Seyfert 2 galaxies) in the low redshift range. Considering both the high redshift and the low redshift samples, 204 galaxies are classified as Narrow Line AGN: 126 of them (62\%) are placed by PCA+UPF algorithms in the late type galaxies cluster, while 47 (23\%) are in the early types cluster and 31 (15\%) are in the green valley region. If we restrict our analysis to the low redshift sample, 95 active galaxies are classified as LINERs: 54 of them (57\%) are in the late types cluster, 22 (23\%) are in the early types one and 19 (20\%) are in the green valley. Conversely, the 30 pure Seyfert 2 galaxies are placed by our PCA+UPF algorithms as follows: 15 of them (50\%) in the late types cluster, 11 (37\%) in the early types region and only 4 (13\%) in the green valley. Though we are facing small number statistics, it is clear that the majority of the analysed type-2 AGN are hosted by galaxies which belong to the blue, late-type cluster. This is quite expected, since our active galaxies span the low luminosity regime, as indicated by the [\ion{O}{iii}]$\lambda$5007 \AA\ line luminosity $10^{5.5} L_{\odot} < L[\ion{O}{iii}] < 10^{9.1} L_{\odot}$ \citep[][]{Bongiorno}.

We also explored the fraction of the selected active nuclei in the various clusters as defined by PCA+UFP with respect to the parent population of all galaxies.
While the fraction of type-2 AGN in each main cluster is around 2\%, this class of objects constitutes $\sim 4\%$ of the galaxies in the PCA+UFP green valley region. At low redshifts, LINERs represent 2\% of the objects in the late type cluster and 3\% of galaxies in the early type one, but they make up 6\% of the green valley galaxies. This picture suggests a possible enhancement of type-2 AGN in the green valley region.
However, since numbers are small -- and therefore errors are large -- this might not be statistically significant. In fact, the observerd type-2 AGN fractions in these subclasses are still compatible with being flat sub-samples extracted purely randomly from the parent sample.

\section{Summary and conclusions}\label{sec:conclusioni}

The classification cube method \citep{Mignoli09} has been extended and applied to the high redshift sample of the zCOSMOS-bright 10k release, exploiting bimodalites in spectral ($D4000$ and $\ion{O}{ii}$ equivalent width), photometric ($B-\mathrm{z}$ colour) and morphological (ZEST classification scheme) properties of the galaxies. In order to overcome some of its limitations (rigidity of the scheme due to its ``hard partitioning'' possibility and nature of misclassifications, reliance on a particular set of data and the difficulty to adopt different variables, a certain degree of arbitrariety in the boundary definitions for the subclassifications) in this work we set up a different classification method based on statistical approaches like the Principal Component Analysis and the Unsupervised Fuzzy Partition (PCA+UFP), that exploits the bimodal nature of galaxy properties in a more organic and rigorous way.

The PCA+UFP analysis is a very powerful and robust tool to probe the nature and the evolution of galaxies in a survey. It allows to define with less uncertainties the classification of galaxies, 
adding the flexibility to be adapted to different parameters: being a fuzzy classification it avoids the problems related to a hard classification. The PCA+UFP method can be easily applied to different datasets: it does not rely on the nature of the data and for this reason it can be successfully employed with others observables (magnitudes, colours) or derived properties (masses, luminosities, SFRs, etc.). 

The agreement between the two classification cluster definitions is very high. ``Early'' and ``late'' type galaxies are well defined by the spectral, photometric and morphological properties, both considering them in a separate way and then combining the classifications (classification cube) and treating them as a whole (PCA+UFP cluster analysis). Differences arise in the definition of outliers: the classification cube is much more sensitive to single measurement errors or misclassifications in one property than the PCA+UFP cluster analysis, in which possible measurement errors are ``averaged out'' during the process.

The PCA+UFP analysis has been applied also to the low redshift sample, substituting $D4000$ and $EW_0[\ion{O}{ii}]$ with $EW_0(\mathrm{H}\alpha$). PCA+UFP analyses, for the high and the low redshift samples, allowed us to behold the \emph{downsizing} effect taking place in the PC spaces: the migration from the blue cloud towards the red clump happens at higher redshifts for galaxies of larger mass. The determination of $M_{\mathrm{cross}}$, the transition mass, is in good agreement with other values in literature.

The green valley objects, as defined with the PCA+UFP cluster analysis, represent also a more coherent sample with respect to classical colour definitions, having the same overall physical properties. Subsequent X-ray and radio analyses could help unveil more the nature of these transitional objects. 

\bibliographystyle{aa} 
\bibliography{mieibib}

\end{document}